\documentclass[notitlepage,tightenlines,aps,pra,reprint,twocolumn,amsmath,amssymb,citeautoscript,nofootinbib,floatfix, longbibliography]{revtex4-1}
\usepackage{float}
\usepackage{graphicx}
\usepackage{dcolumn}
\usepackage{bm}
\usepackage{latexsym,epsfig}
\usepackage{graphicx}
\usepackage{verbatim}
\usepackage{comment}
\usepackage{amsmath} 
\usepackage{amssymb}
\usepackage{stmaryrd}
\usepackage{color}
\usepackage{soul}
\usepackage{epstopdf}
\usepackage{grffile}
\usepackage{mathtools}
\usepackage{physics}
\usepackage[normalem]{ulem}
\usepackage{hyperref}
\hypersetup{
 colorlinks=true, 
 urlcolor=blue,
 citecolor=blue,
 linkcolor=blue
}

\DeclareGraphicsExtensions{.eps}
\newcommand{\beq}{\begin{equation}}
\newcommand{\eeq}{\end{equation}}
\newcommand{\bea}{\begin{eqnarray}}
\newcommand{\eea}{\end{eqnarray}}
\newcommand{\ben}{\begin{eqnarray*}}
\newcommand{\een}{\end{eqnarray*}}
\newcommand{\bfig}{\begin{figure}}
\newcommand{\efig}{\end{figure}}

\newcommand{\ra}{\rangle}

\newcommand{\upa}{\uparrow}
\newcommand{\dna}{\downarrow}

\newcommand{\nn}{\nonumber}

\begin{document}
\title{Topological phase transition through tunable nearest-neighbor interactions in a one-dimensional lattice}

\author{Rajashri Parida$^{1,2}$}
\author{Diptiman Sen$^{3}$}
\author{Tapan Mishra$^{1,2}$}
\email{mishratapan@gmail.com}
\affiliation{
$^1$School of Physical Sciences, National Institute of Science Education and Research, Jatni 752050, India\\
$^2$Homi Bhabha National Institute, Training School Complex, Anushaktinagar, Mumbai 400094, India\\
$^3$Centre for High Energy Physics, Indian Institute of Science, Bengaluru 560012, India}

\date{\today}

\begin{abstract}
We investigate the phase diagram of a one-dimensional model of hardcore bosons or spinless fermions with tunable nearest-neighbor interactions. By introducing alternating repulsive and attractive interactions on consecutive bonds, we show that the system undergoes a 
quantum phase transition from a bond-ordered (BO) phase to a charge-density wave-II (CDW-II) phase as the attractive interaction strength increases at a fixed repulsive interaction. For a specific interaction pattern, the BO phase exhibits topological properties, which vanish when the pattern is altered, leading to a transition from a topological BO phase to a trivial BO phase through a gap-closing point where both interactions vanish. We identify these phases using a combination of order parameters, topological invariants, edge-state analysis and Thouless charge pumping. By extending our analysis beyond half-filling, we explore the phase diagram across all densities and identify the superfluid (SF) and the pair-superfluid (PSF) phases, characterized by single-particle and bound-pair excitations at incommensurate densities. The proposed model is experimentally realizable in platforms such as Rydberg excited or ultracold atoms in optical lattices, offering a versatile framework to study such interplay between topology and interactions in low-dimensional systems. Finally, we present the
complete ground state phase diagram for all possible
signs of the interaction strengths, repulsive or attractive, on consecutive bonds.

\end{abstract}

\maketitle

\section{Introduction}\label{introduction}
The study of topological phases and phase transitions has been a subject of intense research in condensed matter physics~\cite{Hassanreview}. Primarily based on a single-particle picture, the topological phases or topological insulators have shaped our understanding of solid-state physics by introducing a new framework for classifying condensed matter systems. 
The central feature that distinguishes topological phases from non-topological ones is the bulk-edge correspondence, which links the bulk topological invariants to the existence of robust edge or interface states at the boundaries between topologically distinct phases. These edge states are protected by certain underlying symmetries and remain stable against small perturbations until the bulk gap remains finite. Such phases are known as the 
symmetry-protected topological (SPT) phases~\cite{senthil_review_2015,rachel_review}. The robustness of the SPT phases makes them highly promising for applications in fields like topological quantum computing, scattering-free wave transport, and several other advanced technologies~\cite{vonKlitzing2017, senthil_review_2015,Fidkowski,Oshikawa,Xiao-Gang,sptinteract,Pachos_2014,extended_ssh,glide_ssh_ladder,chiral_dynamics}. 

One of the celebrated models that possesses such a non-trivial gapped SPT phase is the one-dimensional Su-Schrieffer-Heeger (SSH) model~\cite{ssh_model} where electrons experience dimerized nearest-neighbor hoppings. The appearance of such unconventional phenomenon in a simple one-dimensional setting has led to numerous exploration of the topological phases in systems described by the SSH model and its variants~\cite{Asboth2016_ssh,smita_ssh_ladder,ssh_3,ssh_su_chen,gloria_platero_ssh,gen_ssh}. Recent progress in the experimental front has propelled the observation of the topological phases in various quantum simulators such as cold atoms in optical lattices, photonic lattices, superconducting circuits, trapped ion arrays, mechanical systems, acoustic systems, electrical circuits, and Rydberg atomic set-ups~\cite{Atala2013,Takahashi2016pumping,Lohse2016,Mukherjee2017,Lu2014,ssh_expt_1,ssh_expt_2, ssh4_expt,Kitagawa2012,Leder2016,browyes,semicoductor_nanolattice,seba_soliton,Liu-superconducting,rydberg_atom_review}.

Topological phases in non-interacting systems have yielded a wealth of insights and established a robust framework based on band topology and topological invariants. However, interacting systems provide a more challenging domain owing to the competing effects of symmetries, strong correlations, lattice topology, and particle statistics~\cite{rachel_review}. 
This has motivated numerous studies which have explored the effects of inter-particle interactions on the stability of topological phases arising from the SSH-type models~\cite{wu_vincet_liu,Taddia2017,Grusdt_topology,manmana_2012,ssh_hubbard_kawakami,hatsugai_prl,DiLiberto2016,DiLiberto2017,pumping_kawakami,fraxanet,ssh_hubbard_ent_entropy,ssh_hubbard_luca,ent_spectrum_ssh_hubbard,Yoshihito_kuno_ssh_hubbard,two_component_luca,juliafare_pumping,hatsugai_prl,zak_phase_graphene,sjia,nersesyan_ssh_ladder,pedro_ssh,dimerized_heisenberg,mohamadi_ssh}. Interestingly, recent studies have shown that strong interactions under proper conditions can favor topological phases ~\cite{haldane,juliafarre_efh,juliafare_quantum,ebh_hidden_order,rossini_ebh,dalla_tore_ebh_prb,Greschner_Mondal,padhan_ladder}.
Even more fascinating is a scenario where interactions alone induce topological phases in systems that are completely trivial in the non-interacting limit~\cite{Mondal_topology,hetenyi_tvvp,rajashri_ladder,harsh_xxz,topo_end_states}.
Recently, several such interaction effects and the interaction-induced SPT phases have been observed using various experimental platforms~\cite{bloch_haldane,suotang_jia_expt,browyes,seba_nphy_pumming,seba_nature_pumping}.

\begin{figure}[t]
\centering
\includegraphics[width=0.8\linewidth]{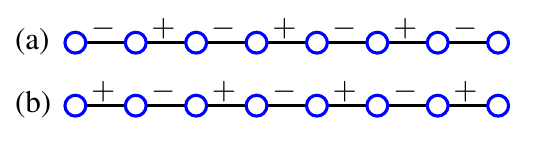}
\caption{Schematic illustration of the interaction pattern in NN bonds of the lattice, where $+$ denotes repulsive interaction and $-$ denotes attractive interaction.} \label{fig:schematic}
\end{figure}

Among the existing experimental platforms the systems of ultracold dipolar quantum gases in optical lattices have emerged as versatile platforms for controlling and manipulating the requisite off-site inter-particle interactions which appropriately mimic the models that host topological phases~\cite{franchesca_ebh,bloch_haldane}. 
By exploiting the tunability of such interactions, we reveal in this work that in one dimensional lattice, an interaction-induced SPT phase can emerge at 
half-filling of hardcore bosons or spinless fermions. By assuming uniform hopping amplitudes of the particles (i.e., no dimerization of the hopping) and allowing attractive and repulsive nearest-neighbor (NN) interactions on the alternate bonds of the lattice as depicted in the schematic in Fig.~\ref{fig:schematic}(a), we show that the system turns topological for any finite values of the interaction strengths, while the non-interacting limit is a gapless superfluid (SF) or Luttinger liquid (LL). This allows for a phase transition from topological to trivial phase by altering the pattern of the NN interaction (i.e., by changing from attractive-repulsive to repulsive-attractive as depicted in the schematic in Fig.~\ref{fig:schematic}(b)) through the point when both the interactions vanish. We demonstrate the signatures of the topological phase and the phase transition through various equilibrium properties. Additionally, we identify further evidence of the topological phase transition via Thouless charge pumping. For completeness, we extend our analysis to explore the situation at all the densities. We also discuss the experimental feasibility of realizing our model. Finally, we present a comprehensive discussion of the ground state phase diagram at half-filling for all possible signs of the interaction strengths.

The structure of this paper is as follows. In Sec.~\ref{modelmethod}, we describe the model under consideration and the methodology used to obtain all the results. Sec.~\ref{results} discusses the key findings for the case where the interaction strengths 
have opposite signs, i.e., repulsive or attractive, on alternate nearest-neighbor bonds. Sec.~\ref{fullphase} discusses the complete ground state phase diagram for all possible signs of the interaction strengths. Finally, Sec.~\ref{conclusion} summarizes the main results. The details of the analytical calculations are
presented in the Appendices.

\section{Model and Method}\label{modelmethod}

We consider a Hamiltonian for a system of interacting hardcore bosons with dimerized NN interactions which is given by
\begin{eqnarray}
 H&=& - t~ \sum\limits_{i}({a}^\dagger_{i}{a}_{i+1}+\text{H.c.}) \nonumber\\
 && + V_1 ~\sum\limits_{i\in\text{odd}}(n_i-\frac{1}{2})(n_{i+1}-\frac{1}{2})
 \nonumber\\
 && + V_2 ~\sum\limits_{i\in\text{even}}(n_i-\frac{1}{2})(n_{i+1}-\frac{1}{2}),
 \label{eq:ham}
\end{eqnarray}
where $i$ denotes the site index, $a_i$ ($a^\dagger_i$) are the annihilation (creation) operators at site $i$, and $n_i =a_i^\dagger a_i$ represents the number operator at site $i$. The parameter $t$ corresponds to the hopping amplitude between NN sites, while $V_1$ and $V_2$ denote the strengths of the dimerized NN interactions. In our numerical calculations presented below,
we set $t=1$ which sets the energy scale in the system. The specific form of the NN interaction terms ensures particle-hole symmetry in the Hamiltonian. 

The Hamiltonian given above can be transformed into a spin Hamiltonian by using the Holstein–Primakoff transformation~\cite{hp_transformation}. This is defined in terms of spin-1/2 operators as
\bea
S_i^+ &=& a_i^\dagger\sqrt{1-a_i^\dagger a_i} ,~~~~S_i^-=\sqrt{1-a_i^\dagger a_i}~a_i, \nn \\
S_i^z &=& a_i^\dagger a_i-\frac{1}{2}, \label{hp_transform}
\eea
where the bosonic occupation state $|0\rangle$ and $|1\rangle$ map to the spin states $|\downarrow\rangle$ and $|\uparrow\rangle$ respectively. The Hamiltonian for the spin chain can then be written as 
\begin{eqnarray}
 H&=& - 2t ~\sum\limits_{i}(S_i^xS_{i+1}^x+S_i^yS_{i+1}^y) \nonumber \\
 &&+ V_1 ~\sum\limits_{i\in\text{odd}} S_i^zS_{i+1}^z+V_2 ~\sum
 \limits_{i\in\text{even}} S_i^zS_{i+1}^z. \label{eq:spin_ham}
\end{eqnarray}

The Hamiltonian in Eq.~(\ref{eq:spin_ham}) can also be mapped to a model of interacting spinless fermions using the Jordan-Wigner transformation~\cite{jw_transform} defined as
\bea
S_j^- &=& e^{i\pi\sum\limits_{k=1}^{j-1}c_k^\dagger c_k}c_j,~~~~ S_j^+=c_j^\dagger 
e^{-i\pi\sum\limits_{k=1}^{j-1}c_k^\dagger c_k}, \nn \\
S_j^z &=& c_j^\dagger c_j ~-~ \frac{1}{2} ~=~ n_i^f ~-~ \frac{1}{2}.
\eea
This mapping leads to a spinless fermionic Hamiltonian which can be written as
\begin{eqnarray}
 H&=& - t ~\sum\limits_{i}({c}^\dagger_{i}{c}_{i+1}+\text{H.c.}) \nonumber\\
 && + V_1 ~\sum\limits_{i\in\text{odd}}(n_i^f-\frac{1}{2})(n_{i+1}^f-\frac{1}{2}) \nonumber\\
&& + V_2 ~\sum\limits_{i\in\text{even}}(n_i^f-\frac{1}{2})(n_{i+1}^f-\frac{1}{2}).
\label{eq:fermionic_ham}
\end{eqnarray}
We have assumed open boundary conditions for this mapping, as will be used throughout this work. With periodic boundary conditions, however, an additional boundary term arises, which depends on the total number of fermions (or total $S_z$) in the chain. The above mappings ensure that the models shown in Eqs.~(\ref{eq:ham}), (\ref{eq:spin_ham}) and (\ref{eq:fermionic_ham}) share similar ground state phase diagrams. The results presented below are all obtained by considering the model shown in Eq.~(\ref{eq:ham}) which corresponds to a system of hardcore bosons. 

The above model has been previously studied using mean-field theory~\cite{fisher_mean_field,hrk_mean_field,p_zoller_mean_field,stoof_mean_field}, the
density-matrix renormalization group (DMRG)~\cite{white1992,schollowck_dmrg_rev} method as well as the bosonization approach, for both hardcore bosons and spins ~\cite{Mondal_topology,harsh_xxz}. These findings highlight that while the 
non-interacting version of the model remains gapless, introducing a dimerized NN repulsion for hardcore bosons (or dimerized repulsive $ZZ$ exchange interaction for spins) leads to a topological phase and an associated topological phase transition. The topological phase transition occurs
as a function of the interaction dimerization which is tuned by fixing the NN interaction strengths for the even (odd) bonds and varying it on the odd (even) bonds of the lattice. This phase transition is marked by a gap-closing point, occurring when the interactions are of equal strengths. However, the scenario considered in this work is completely different as we assume the NN interaction to be attractive in nature in either all the even or odd bonds (see Fig.~\ref{fig:schematic}). 

We explore the ground state properties of the system using the DMRG method~\cite{Verstraete_rev,schollowck_mps} under open boundary condition (OBC) and for specific cases we use the exact diagonalization (ED) approach with periodic boundary condition (PBC). For the DMRG simulations, we consider a maximum system size of $L=200$ and set the maximum bond dimension to $1000$ to minimize truncation errors. On the other hand, $L=16$ is used for the ED simulations. Unless stated otherwise, all quantities are extrapolated to the thermodynamic limit ($L\rightarrow\infty$) to minimize finite-size effects. We also complement some of our numerical findings through analytical arguments.

\section{Results}\label{results}

In this section, we discuss our findings in detail. We first present the bulk phase diagram and then analyze the topological properties at half-filling. We also provide the signature of the topological phase through Thouless charge pumping. Subsequently, we analyze the scenario by moving away from half-filling following which we discuss the experimental feasibility of the model.
In the end, we briefly discuss the full ground state phase diagram by considering all possible signs of the interaction strengths - both attractive and repulsive.

\begin{figure}[t]
\centering
\includegraphics[width=1.\linewidth]{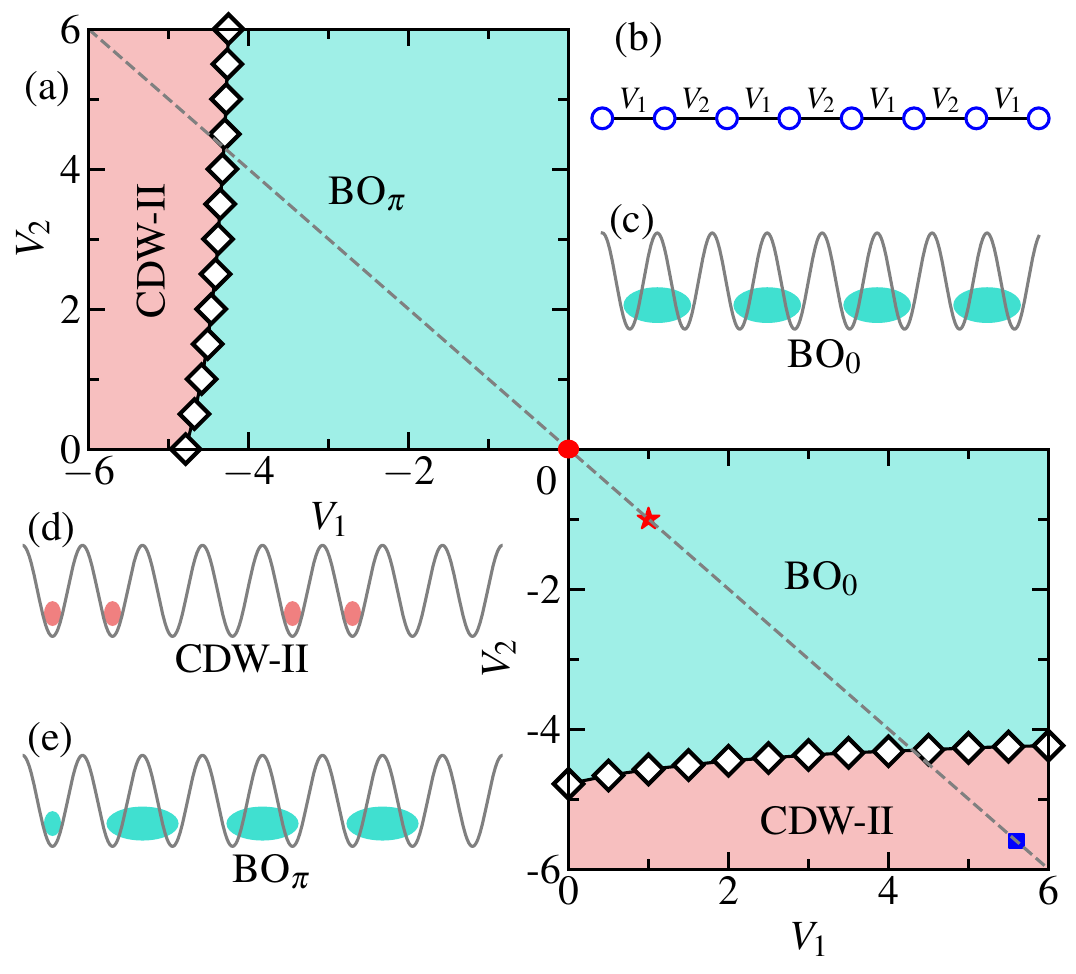}
\caption{(a) The phase diagram of the model in the $V_1-V_2$ plane at half-filling shows the BO$_0$, BO$_\pi$ and the CDW-II phases. The red circle at the origin depicts the gapless critical point. The diamond symbols indicate the boundary separating the BO and the CDW-II phases. (b) shows a cartoon diagram depicting the interaction pattern considered. (c), (d) and (e) show the schematics for the BO$_0$, CDW-II, and BO$_\pi$ phases, respectively.} \label{fig:phasedia}
\end{figure}

\subsection{Phase diagram at half-filling}

The ground state phase diagram of the system of hardcore bosons at half-filling is depicted in Fig.~\ref{fig:phasedia}(a) in the $V_1$-$V_2$ plane. The upper left (lower right) corner of the figure corresponds to the situation when $V_1 < 0$, $V_2 > 0$ ($V_1 > 0$, $V_2 < 0$). For each case, we numerically find 
that the entire phase diagram is gapped and there is a gapped-to-gapped transition. In the following, we identify the two gapped phases as the bond order (BO) and the charge-density wave (CDW) phases except at the origin where the system is gapless (denoted 
by a red circle). 

\begin{figure}[t]
\centering
\includegraphics[width=0.7\linewidth]{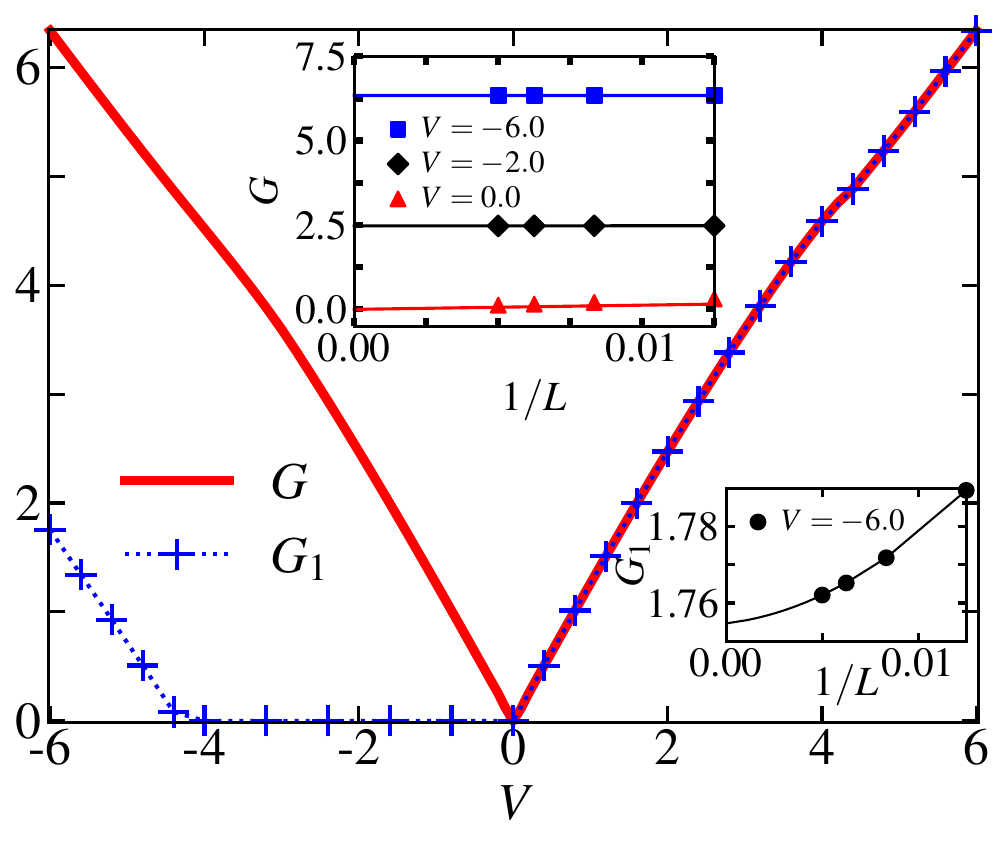}
\caption{Single-particle excitation gap $G_1$ and the bulk gap $G$ are plotted as a function of $V$ along the dashed line drawn in Fig.~\ref{fig:phasedia}. $G$ (solid curve) remains finite for all the values of $V\neq 0$. $G_1$ vanishes in the region $-4.3 \lesssim V \leq 0$ and remains finite else where. 
The insets show the finite-size extrapolation of gaps $G_1$ (lower panel) and $G$ (upper panel) for three exemplary parameter values corresponding to the gapless ($V=0$) and the gapped phases ($V=-2.0$ and $V=-6.0$). In the inset, the solid lines are the fitted curves and the symbols represent the numerical data.
} \label{fig:gap}
\end{figure}

We find that for any finite value of $V_2$ when $V_1=0$, the ground state is a BO phase (the upper left part of the phase diagram). However, as $V_1$ is increased on the attractive side, a transition to a CDW phase occurs. This results in a 
gapped-to-gapped transition as a function of attractive $V_1$ for each value of repulsive $V_2$. A similar situation is found upon interchanging the nature of the interactions i.e., by making $V_1$ repulsive and $V_2$ attractive; the corresponding phase diagram is shown in the bottom right corner of Fig.~\ref{fig:phasedia}(a). In the following, we quantify these phases by computing different relevant physical quantities as a function of $V_1=-V_2=V$ chosen for simplicity. This corresponds to an exemplary cut drawn in the phase diagram which is denoted by a dashed line in Fig.~\ref{fig:phasedia}(a). Along this line, when $V<0$, all the odd bonds experience attractive interaction while the even bonds experience repulsive interaction due to the boundary condition chosen (i.e. OBC). Conversely, for $V>0$, the odd bonds become repulsive, and even bonds become attractive.
\begin{figure}[b]
\centering
\includegraphics[width=0.8\linewidth]{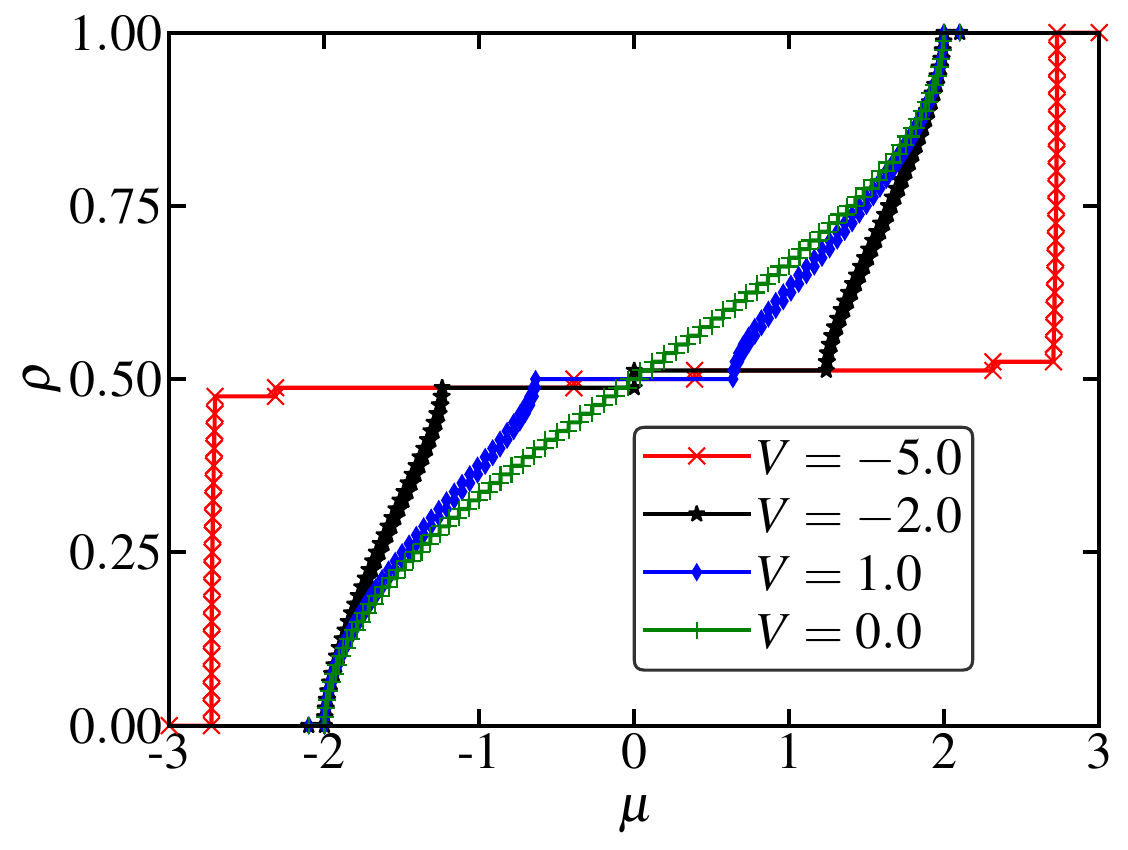}
\caption{The change in density ($\rho$) as a function of change in chemical potential ($\mu$) for different values of $V$ for a system size $L=80$. The plateaus at half-filling ($\rho=1/2$) indicate the gap in the system and the length of the plateaus give the estimate of the gaps.} \label{fig:rho_mu_gap}
\end{figure}
\begin{figure}[t]
\centering
\includegraphics[width=0.8\linewidth]{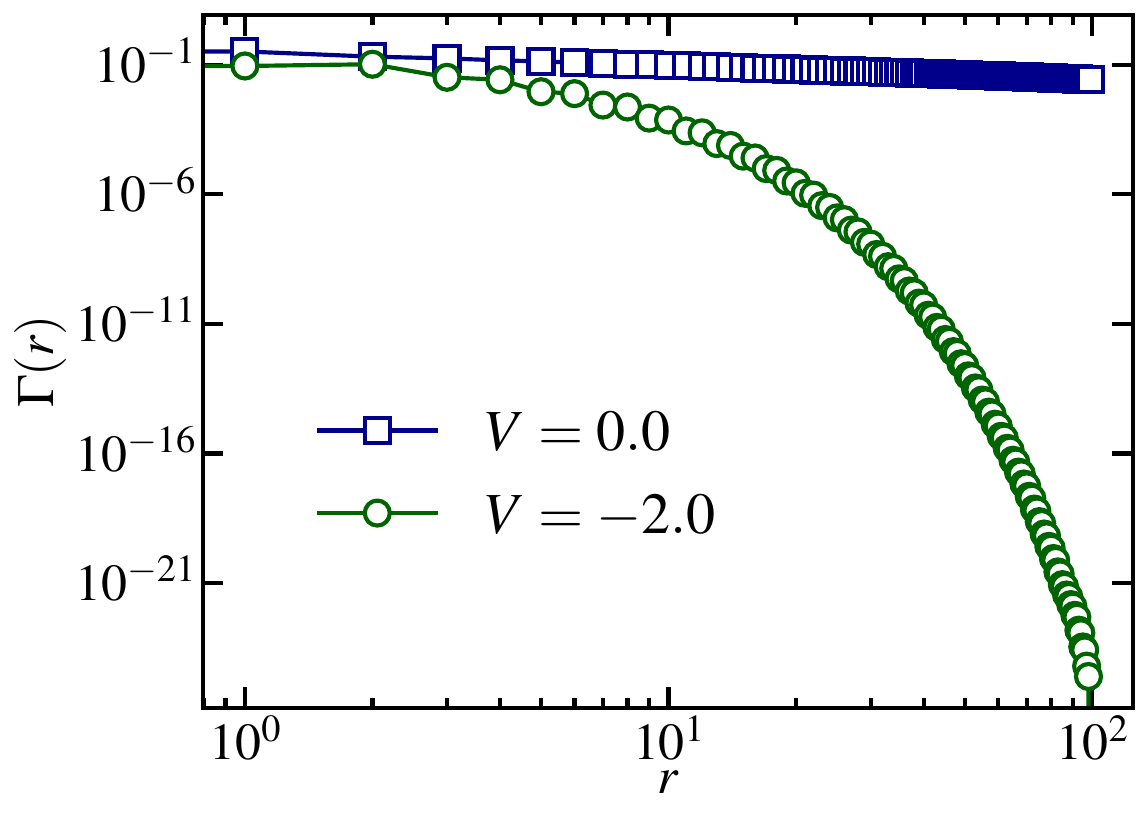}
\caption{The correlation function $\Gamma(r)$ plotted as a function of distance 
($r=|i-j|$) for a system size $L=200$ showing power-law and exponential decays for $V=0.0$ and $V=-2.0$, respectively. The correlation values taken from $L/4$ to $3L/4$ to avoid the edge effects.} \label{fig:correlation_decay}
\end{figure}
\begin{figure}[b]
\centering
\includegraphics[width=1.\linewidth]{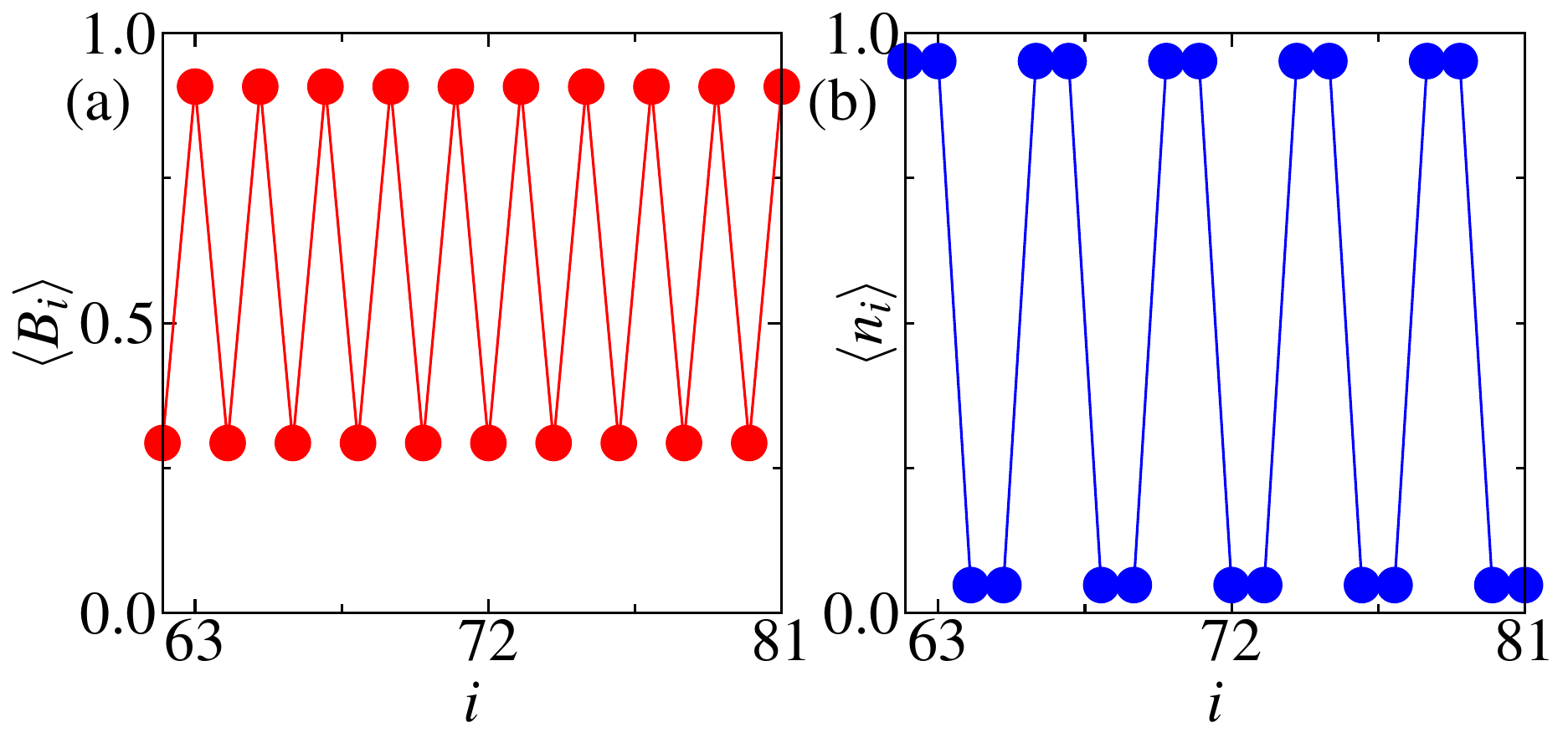}
\caption{(a) and (b) display the bond energy $\langle B_i\rangle$ and the on-site particle density $\langle n_i \rangle$ as a function of $i$ for $V=1.0$ and $V=5.6$ marked as red star and blue square, respectively in the phase diagram shown in Fig.~\ref{fig:phasedia}. The finite oscillations in the
bond energy for $V=1.0$ signify the presence of the bond ordering in the system, while the finite on-site density oscillations indicate the presence of a CDW ordering exhibiting a density distribution of $\cdots 11001100 \cdots$ type.} \label{fig:bo_oscillation}
\end{figure}

\begin{figure}[t]
\centering
\includegraphics[width=0.8\linewidth]{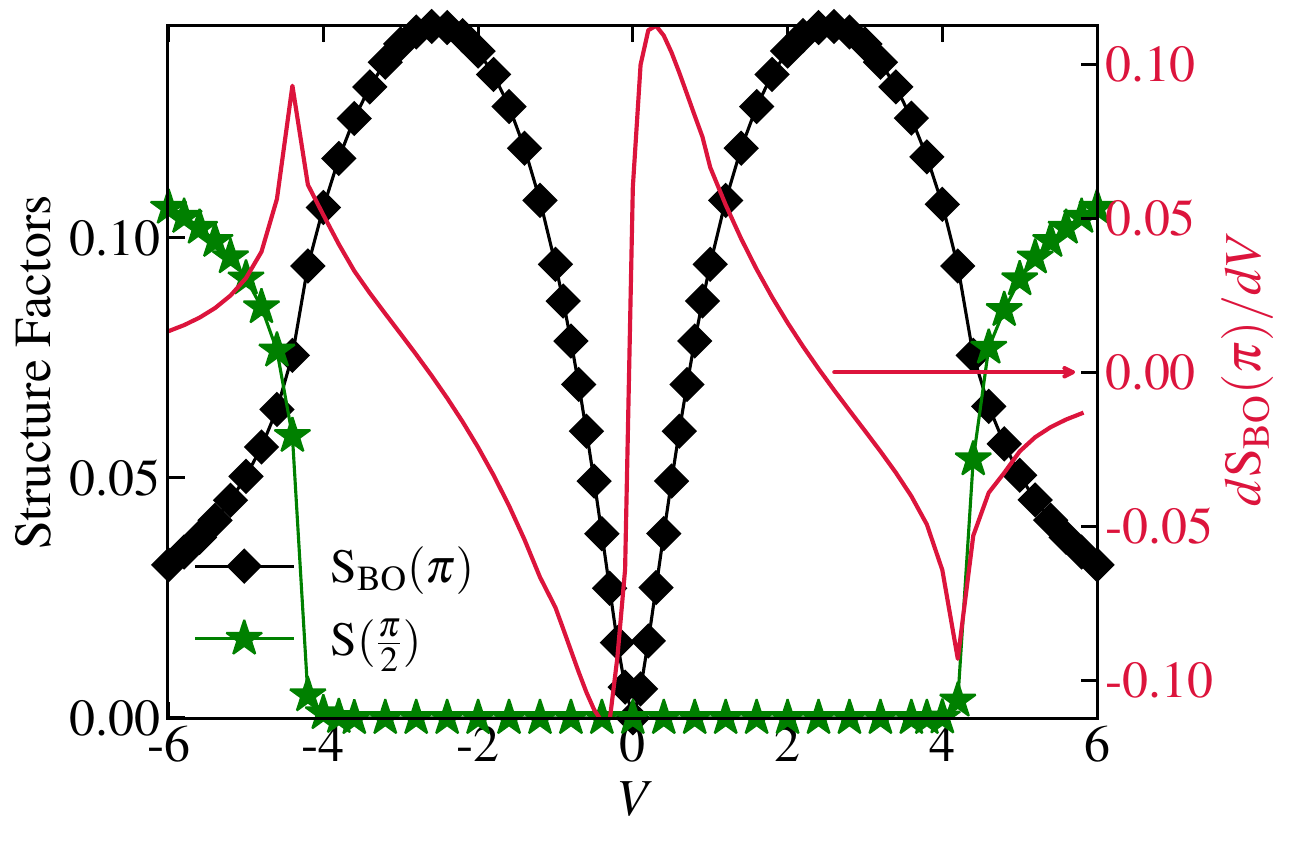}
\caption{Extrapolated values of S$_{\text{BO}}(\pi)$ (black diamonds) and S$(\pi/2)$ (green stars) as a function of $V$ computed by taking a maximum system size of $L=200$. The derivative $d\text{S}_{\text{BO}}({\pi})/dV$ (red solid curve) is plotted to show that S$_{\text{BO}}(\pi)$ also has non-analytic behavior at the BO to CDW-II phase transition point.}
\label{fig:structure_factor}
\end{figure}

\begin{figure*}[t]
\centering
\includegraphics[width=0.8\linewidth]{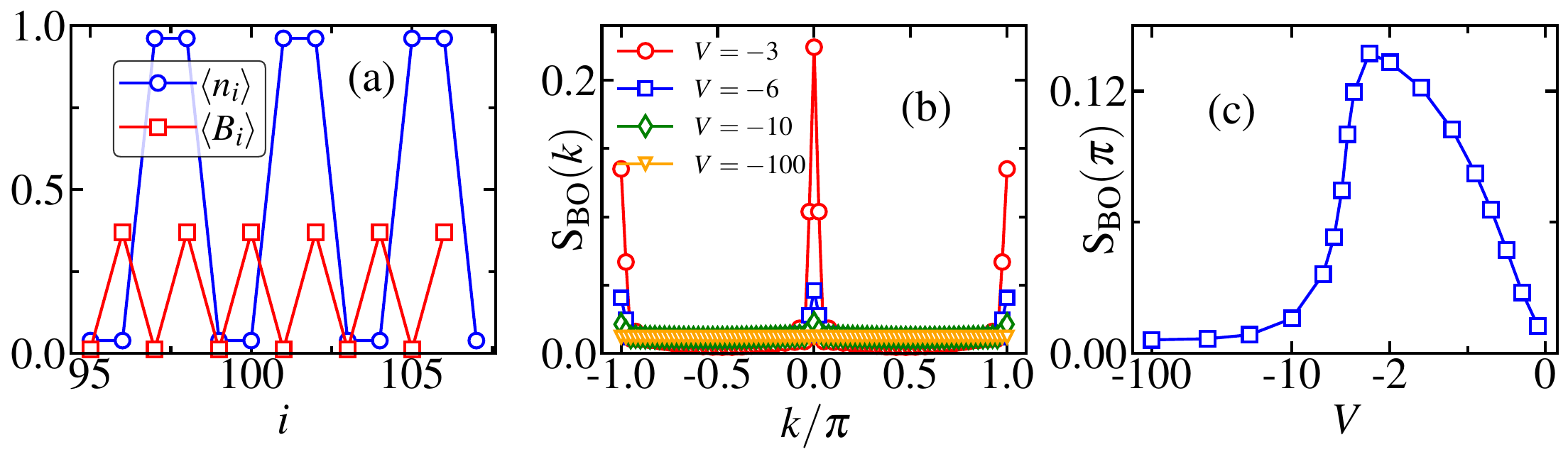}
\caption{(a) The onsite particle density $\langle n_i \rangle$ (blue circles) and the bond energy $\langle B_i \rangle$ (red squares) as functions of $i$ for $V=-6$ which corresponds to the CDW-II phase. Results are shown for a system of size $L=200$ under OBC, with the central region of the lattice displayed for better visibility. (b) Bond-order structure factor S$_{\text{BO}}(k)$ for system size $L=80$ at various interaction strengths $V = -3$ (red circles), $-6$ (blue squares), $-10$ (green diamonds), and $-100$ (orange triangles). (c) Shows the peak value S$_{\text{BO}}(\pi)$ as a function of $V$. $x$-axis is shown in log-scale for better visibility.}
\label{fig:large_int_bo_sf}
\end{figure*}

To extract the gapped phases in the phase diagram we rely on the particle excitation gaps defined as
\begin{equation} G_1=\mu^+_1-\mu^-_1, \end{equation}
where
\begin{equation} \mu^+_1=E_{N+1}-E_N ~~~\text{and}~~~ \mu^-_1=E_N-E_{N-1},
\label{eq:gap} \end{equation}
are the chemical potentials corresponding to adding and removing a particle respectively in the system, and $E_N$ represents the ground state energy of the system with $N$ bosons. In Fig.~\ref{fig:gap}, we plot the extrapolated values of the gap $G_1$ (dotted plus curve) as a function of $V_1=-V_2=V$. It can be seen that the gap remains finite in the region $V>0$. However, in the region $V\leq 0$, the gap $G_1$ vanishes up to a critical $V\sim -4.3$, after which it becomes finite. In the following, we will show that while the vanishing of $G_1$ at $V=0$ is due to the gapless nature of the system at that point, for $-4.3\lesssim V<0$, the zero gap is due to the topological nature of the system. In order to show that the system is gapped in all the other parameter regimes except $V=0$, we
examine the change in density ($\rho$) as a function of the chemical potential ($\mu$) ~\cite{rho_mu_furusaki,rho_mu_tapan,rho_mu_sebastian} for different values of $V$ which is shown in Fig.~\ref{fig:rho_mu_gap}. The density is defined as $\rho=N/L$, where $N$ is the total particle number and $L$ is the system size. The gapped nature can be seen as the plateaus in the $\rho-\mu$ curve at half-filling and the length of the plateau represents the gap ($G$) in the system. We obtain the gap as
\begin{eqnarray} G=\mu^+-\mu^-, 
\end{eqnarray}
where $\mu^+$ and $\mu^-$ are the chemical potentials corresponding to the right and left ends of the plateau for each curve. The extrapolated values of $G$ (red solid curve) as a function of $V$ are plotted in Fig.~\ref{fig:gap} where the finite values of $G$ for $V\neq0$ confirm the gapped nature of the phases.
 At the gapless point $V=0$, the system exhibits off-diagonal quasi-long-range order which is reflected in the power-law decay of the correlation function 
\begin{eqnarray} 
\Gamma(r=|i-j|)=\langle a_i^\dagger a_j \rangle
\label{eq:sf_corr}
\end{eqnarray}
and is plotted in Fig.~\ref{fig:correlation_decay} (blue squares). This is a signature of the SF phase in one dimension.
For comparison, we also show the exponential decay of $\Gamma(r)$ for $V=-2.0$ (green circles) for which the system is gapped. 

We now turn to quantify the gapped phases. As mentioned already, these are the BO and the CDW phases. The BO phase is characterized by finite dimerization of the particles in the NN bonds. To quantify this feature, we use the bond kinetic energy $\langle B_i \rangle$, where $B_i = a_i^\dagger a_{i+1} + \text{H.c.}$, exhibits finite oscillations as a function of the bond index $i$, as shown in Fig.~\ref{fig:bo_oscillation} (a) for $V=1.0$ (the point marked as a red star in Fig.~\ref{fig:phasedia}). At the same time, the CDW order manifests itself as finite oscillations in the real-space density $\langle n_i\rangle$ which has the form $\cdots 11001100 \cdots$ as depicted in Fig.~\ref{fig:bo_oscillation} (b) for $V=5.6$ (the point marked as 
a blue square in Fig.~\ref{fig:phasedia}). This pattern of the density distribution is the signature of a CDW-II phase.

A concrete signature of the BO phase and the CDW-II phase is the presence of finite peaks in their respective structure factors, defined as
\begin{equation}
\text{S}_{\text{BO}}(k)=\frac{1}{L^2}\sum_{i,j}e^{ik|i-j|}\langle B_iB_j \rangle,
\label{eq:bo_stfc}
\end{equation}
and 
\begin{equation}
\text{S}(k)=\frac{1}{L^2}\sum_{i,j}e^{ik|i-j|}\langle n_in_j\rangle,
\label{eq:cdw_stfc}
\end{equation}
where, S$_{\text{BO}}(k)$ and S$(k)$ are the bond order and the charge-density wave structure factors, respectively. To minimize the boundary effects, we consider bond and density correlation functions in the range from $L/4$ to $3L/4$ throughout our calculations. The extrapolated values of S$_{\text{BO}}(\pi)$ and S$(\pi/2)$ are shown in Fig.~\ref{fig:structure_factor} as black diamonds and green stars respectively as a function of $V$ (along the dashed line in Fig.~\ref{fig:phasedia}). The finite values of S$(\pi/2)$ for $V\lesssim-4.3$ and $V\gtrsim4.3$ clearly indicates the CDW-II regions. At the same time, we find that the S$_{\text{BO}}(\pi)$ is finite for all the values of $V$ except at $V=0$ where the system is in the SF phase. 
Though S$_\text{BO}(\pi)$ is finite for $|V|\gtrsim 4.3$, the behaviour of S$_{\text{BO}}(\pi)$ near $|V|\sim 4.3$ is non-analytic, which becomes evident upon examining its derivative, $d\text{S}_{\text{BO}}(\pi)/dV$, shown as red solid lines. The clear discontinuity in $d\text{S}_{\text{BO}}(\pi)/dV$, and the sharp increase in S$(\pi/2)$ near $|V|\sim 4.3$ confirm the transition into the CDW-II phase. In the CDW-II phase, the bond-order structure factor S$_{\text{BO}}(\pi)$ remains finite due to the combined effects of the finite hopping term and the specific occupation pattern of the CDW-II phase. As shown in Fig.~\ref{fig:large_int_bo_sf}(a), bonds between sites with occupations $11$ or $00$ exhibit lower bond energy, while those between $10$ or $01$ show higher bond energy, because of the finite contribution from the hopping term. This leads to a finite bond modulation that gradually vanishes in the strong interaction limit, where hopping becomes negligible. To further clarify the behavior of S$_\text{BO}(\pi)$, we plot S$_\text{BO}(k)$ as a function of the wave vector $k$ for various interaction strengths in Fig.~\ref{fig:large_int_bo_sf}(b). A pronounced peak at $k=\pi$ (because of the oscillatory bond energy pattern shown using red squares in Fig.~\ref{fig:large_int_bo_sf}(a)) is observed for smaller interaction strengths, but it disappears completely at $V=-100$ (orange triangles). The evolution of the peak height with increasing interaction strength is shown in Fig.~\ref{fig:large_int_bo_sf}(c), where it vanishes in the large interaction limit.

\begin{figure}[b]
\centering
\includegraphics[width=1.0\linewidth]{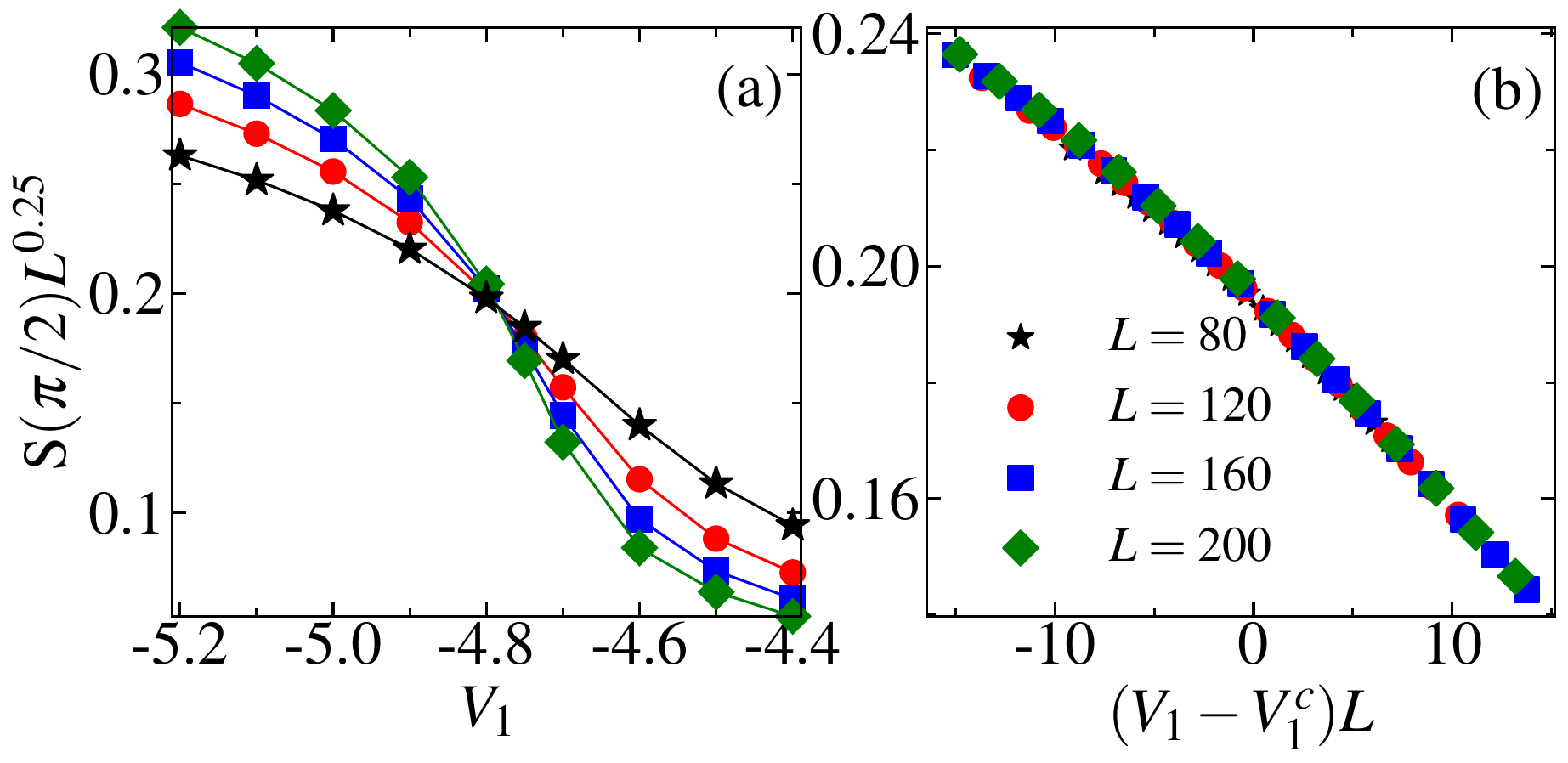}
\caption{(a) Finite-size scaling of S$(\pi/2)$ according to Eq.~\ref{eq:critical_scaling} for various values of $V_1$ at fixed $V_2=0$. The intersection of the curves marks the critical point for the BO to CDW-II phase transition. (b) The scaled S$(\pi/2)$ is plotted against the scaled $V_1$, showing a data collapse onto a single curve, confirming $V_1^c = -4.78$ as the transition point.}
\label{fig:collapse_sec_quad}
\end{figure}

To accurately determine the phase transition boundary between the BO and the CDW-II phases, we perform a finite-size scaling of the CDW-II structure factor S$(\pi/2)$ by varying $V_1$ for fixed $V_2$ values. 
Our results show that the BO to CDW-II phase transition belongs to the quantum Ising universality class. The critical point is identified through the appropriate scaling behavior of the CDW-II structure factor as 
\begin{eqnarray}
 \text{S}(\pi/2)L^{2\beta/\nu} = F[(V_1-V_1^c)L^{1/\nu}],
\label{eq:critical_scaling}
\end{eqnarray}
where, $\beta$ and $\nu$ are the magnetization and correlation length critical exponents~\cite{Mondal_topology,mishra_rigol}. We plot S$(\pi/2)L^{2\beta/\nu}$ as a function of $V_1$ for $V_2=0$ in Fig.~\ref{fig:collapse_sec_quad} (a) with $\beta=1/8$ and $\nu=1$. The curves for different system sizes are found to cross at the critical point $V_1^c\sim-4.78$. Moreover, a perfect data collapse for various system sizes is achieved using the scaling function in Eq.~(\ref{eq:critical_scaling}) at $V_1^c=-4.78$ as shown in Fig.~\ref{fig:collapse_sec_quad} (b) which confirms the critical point. This approach allows us to map out the entire phase boundary as presented in the upper left part of Fig.~\ref{fig:phasedia} (a) (shown using diamond symbols). A similar situation is found upon varying $V_2$ for each values of $V_1$ resulting in the phase boundary shown in the lower right part of Fig.~\ref{fig:phasedia} (a). 

The upper left (lower right) part of the phase diagram in Fig.~\ref{fig:phasedia} (a) shows that as the value of $V_2$ ($V_1)$ is increased, the critical point of the BO to CDW-II transition slowly shifts towards smaller values of attractive $V_1$ ($V_2$). Moreover, the transition boundary seems to saturate at a particular value of $V_1$ ($V_2$) in the limit of large $V_2$ ($V_1$). 
The boundary in the limit of large $V_2$ (or large $V_1$) can be found using second-order perturbation theory, as shown in Appendix~\ref{app:appen_degenpert1}.
For example we find that for $V_2 \gg t$ and $|V_1|$, the location of the boundary between the BO 
and the CDW-II phases in upper left part of Fig.~\ref{fig:phasedia} is given as
\beq V_1 ~=~ -~ 4 t ~-~ \frac{2t^2}{V_2}. \label{v1v2} \eeq
This formula clearly suggests that the critical point saturates to $V_1=-4.0$. A similar analysis of the phase transition in the lower right part of Fig.~\ref{fig:phasedia} reveals that the critical point for the BO to CDW-II transition saturates to $V_2=-4.0$ in the limit of large $V_1$.

From the analysis above, it is evident that the entire phase diagram comprises of two distinct gapped phases at half-filling, namely the BO phase and the CDW-II phase. Additionally, a point where the system exhibits features of a SF phase appears at $V_1=-V_2=0$ which corresponds to the non-interacting limit of the model. In the following we will show that the BO phase in the regime of $V_1<0$ and $V_2>0$ is topological in nature and the one in the regime of $V_1>0$ and $V_2 < 0$ is trivial. We characterize the nature of these phases in detail in the following subsection by taking the same exemplary cut $V_1=-V_2=V$ as previously mentioned.

\subsection{Topological Properties}

Here we demonstrate that the two BO phases appearing in the phase diagram (Fig.~\ref{fig:phasedia} (a)) on either side of $V_1=-V_2=V=0$ are topologically distinct, and the transition occurring at $V=0$ is a topological phase transition.
\begin{figure}[t]
\centering
\includegraphics[width=0.8\linewidth]{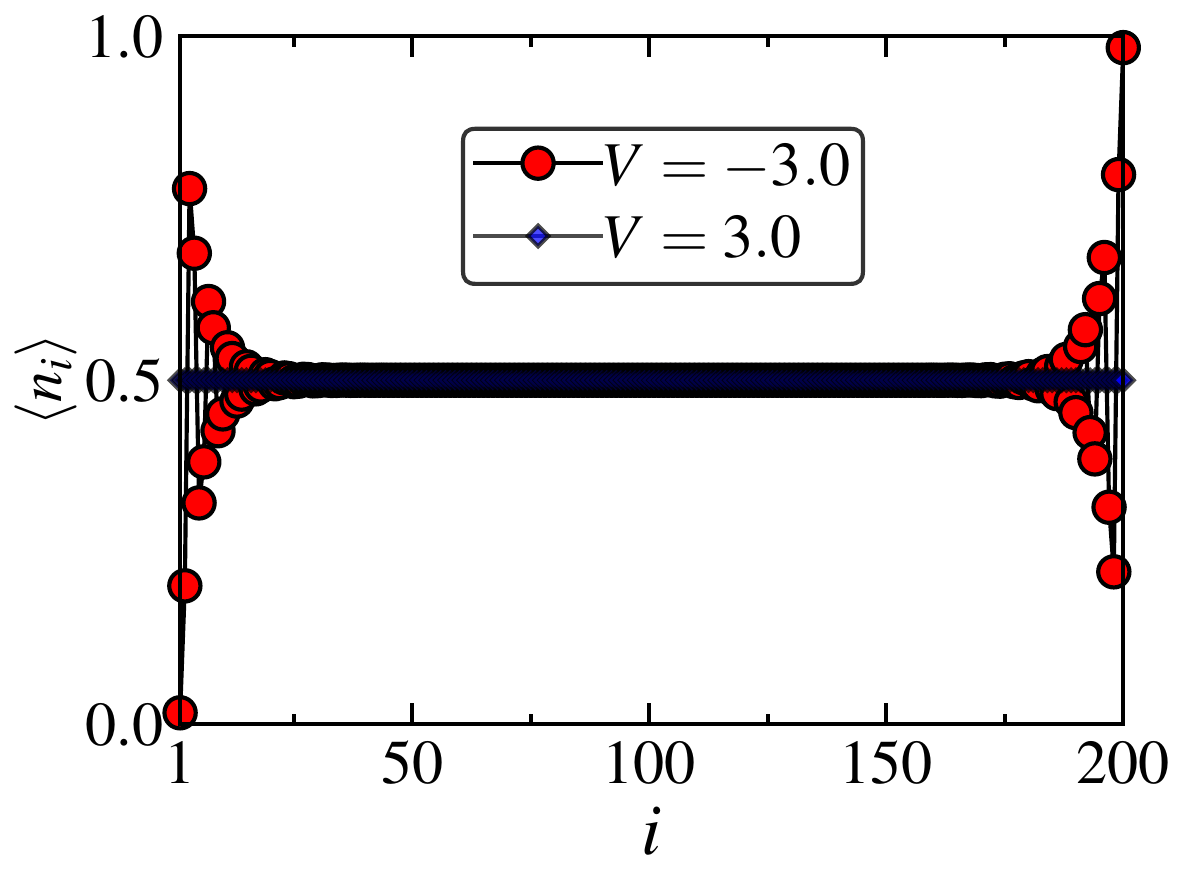}
\caption{On-site particle density $\langle n_i\rangle$ is plotted against the site index $i$ in the BO$_\pi$ and BO$_0$ phases at $V=-3.0$ (red circles) and $V=3.0$ (blue diamonds), respectively for a system size of $L=200$ sites.}
\label{fig:edge_state}
\end{figure}
One of the most significant consequences of the bulk topology is the presence of degenerate edge modes in OBC. Therefore, as an initial step to characterize the topology of the system, we plot the on-site particle density $\langle n_i\rangle$ as a function of the
site index for two different BO regimes. Fig.~\ref{fig:edge_state} shows the plot of $\langle n_i \rangle$ as a function of $i$ for $V=-3.0$ (red circles) and $V=3.0$ (blue diamonds) for a system size $L=200$. The larger values of $\langle n_i \rangle$ near the edge sites for $V=-3.0$ compared to uniform density distribution for $V=3.0$ suggests a topological difference between the two types of BO phases.

\begin{figure}[t]
\centering
\includegraphics[width=1.\linewidth]{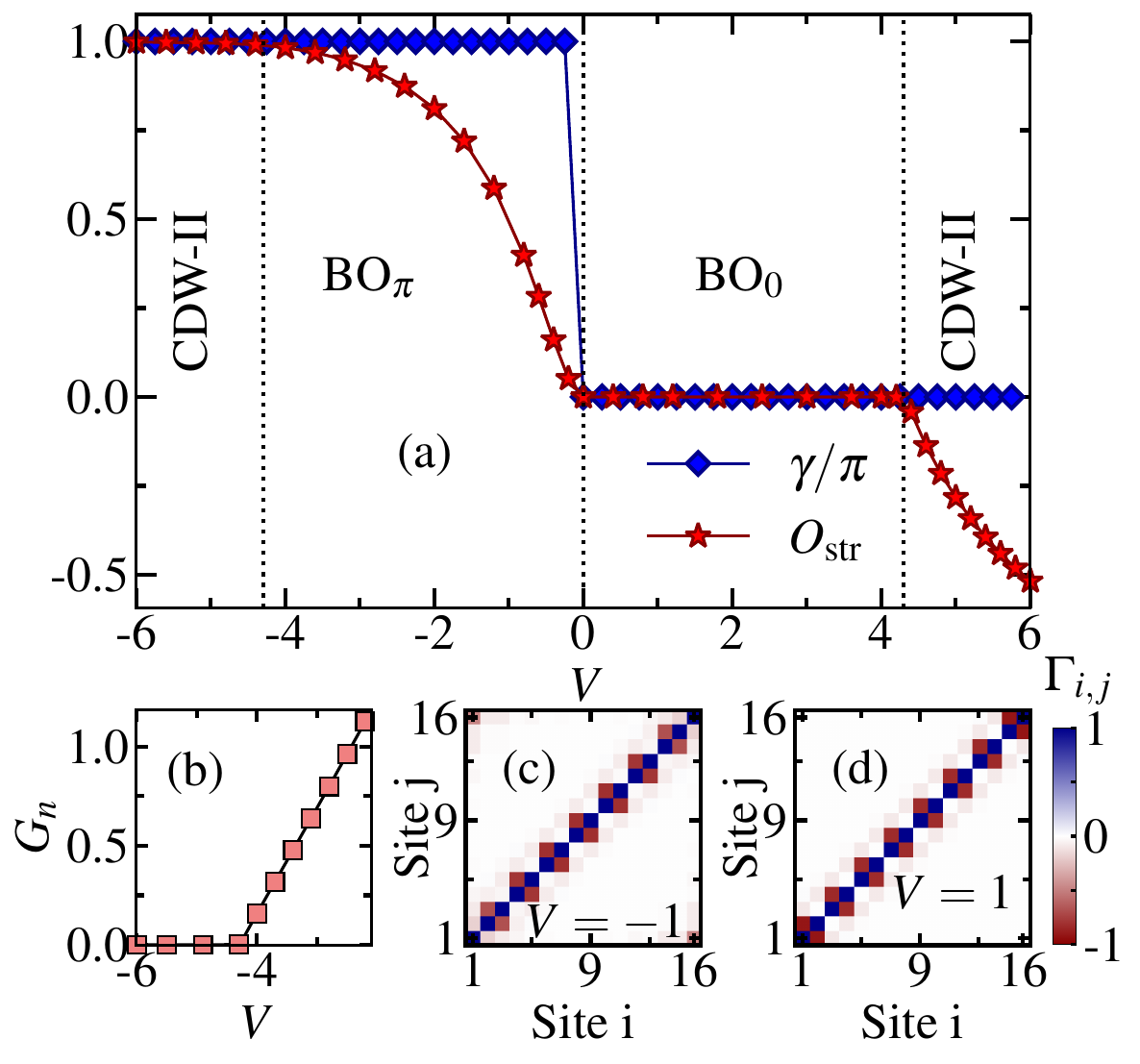}
\caption{(a) Plot of the winding number $\gamma/\pi$ (blue diamonds) and the string order parameter $O_{\text{str}}$ (red stars) as a function of $V$. (b) Plot of the extrapolated value of $G_n$ under periodic boundary conditions, taking a maximum system size of $L=120$ sites. The vertical dashed lines in (a) separate different phases. The density-density correlation matrix $\Gamma$ for $V=-1.0$ and $1.0$ is shown in (c) and (d), respectively, calculated using ED for a system size of $L=16$ sites.}
\label{fig:invariant_correlation_matrix}
\end{figure}

To concretely establish the topological nature of the BO phase appearing in the regime $-4.3\lesssim V<0$ and to characterize the topological phase transition between the two BO phases at $V=0$, we calculate the bulk topological invariant of the system in this regime at various values of $V$ for a system size of $L=16$ using ED. The topological invariant for the interacting system is calculated using twisted boundary conditions, where the hopping strength at the boundary is modified as $t \xrightarrow{} te^{i\theta}$, with $\theta$ being the twist angle~\cite{Zak1989,berry_phase_resta,Grusdt_topology}. As $\theta$ is varied from $0$ to $2\pi$, the ground state wave function $|\psi(\theta)\rangle$ accumulates a phase, known as the Berry phase ($\gamma$)~\cite{berry_phase}, which is determined using the formula
\begin{equation}
\gamma = i\int_0^{2\pi} d \theta ~\langle\psi(\theta)|\partial_\theta|\psi(\theta)\rangle.
\label{eq:berry_phase}
\end{equation}
The Berry phase is $0$ if the bulk of the system is trivial whereas it is $\pi$ if the bulk is topological in nature. We compute $\gamma$ for different values of $V$ and plot $\gamma/\pi$ as a function of 
$V$ in Fig.~\ref{fig:invariant_correlation_matrix}(a) (blue diamonds). The plot shows that $\gamma/\pi$ changes from $1$ to $0$ at 
$V=0$ signifying a phase transition from a topological to a trivial BO phase at $V=0$. Based on the distinct Berry phases, we denote the topological and the trivial BO phases as BO$_\pi$ and BO$_0$ respectively (Fig.~\ref{fig:phasedia} (a)). Additionally, note that the value of $\gamma/\pi=1$ in the CDW-II phase appearing in the region of $V<0$, despite its non-topological nature. However, this phase lacks other topological signatures, such as the presence of degenerate zero-energy edge states as discussed in Fig.~\ref{fig:gap}. We support this conclusion through additional diagnostics discussed below, which collectively confirm the non-topological nature of the CDW-II phase. 

The apparent nontrivial value of $\gamma/\pi$ in the CDW-II phase arises due to the finite-size system under consideration. Note that the CDW-II phase emerges due to the spontaneous inversion symmetry breaking, where the excitation gap remains finite for finite systems and closes only in the thermodynamic limit. However, one can see that the value of $G$ remains finite at the BO$_\pi$ to CDW-II phase transition point (see Fig.~\ref{fig:gap} ). It is important to emphasize that a quantum phase transition requires the excitation gap $G_n=E_1(N)-E_0(N)$ to vanish, where $E_0(N)$ and $E_1(N)$ represent the ground state and first excited state energy, respectively~\cite{Sachdev_qpt_book}. However, in this scenario, $G_n$ vanishes throughout the entire range of $V$  because, the BO$_\pi$ phase is two-fold degenerate under OBC due to the presence of the edge states and the CDW-II phase is also two-fold degenerate. In such a situation, the BO$_\pi$ and the CDW-II phases can not be distinguished from the excitation gap. In contrast, by employing PBC removes these degeneracies and reveals the intrinsic bulk gap of the BO$_\pi$ phase. We plot the extrapolated value of $G_n$ in Fig.~\ref{fig:invariant_correlation_matrix}(b) under PBC which shows that the excitation gap closes at  $V\sim -4.3$, confirming a phase transition from the BO$_\pi$ to the CDW-II phase. Beyond this point ($V\lesssim -4.3$), the gap vanishes in the entire CDW-II phase, consistent with the expected degeneracy of the two $Z_2$-symmetry-broken ground states in the thermodynamic limit.
Since a change in the topological invariant requires gap closing, the Berry phase remains unchanged across the transition in finite-size systems. This robustness of the topological invariant against spontaneous symmetry-breaking has been reported in earlier works~\cite{arpit_tanmoy}, and similar behavior has also been observed in related models~\cite{pedro_ssh}.

Another interesting feature of topological systems is the presence of non-local string correlations~\cite{ebh_hidden_order,dalla_tore_ebh_prb,den_nijs,Tasaki,Hida,string_ladder}, which can be quantified by a string order parameter defined as
\begin{equation}
 O_{\text{str}}(r) ~=~ -~ \langle Z(i)e^{i\frac{\pi}{2}\sum_{j=i+1}^{k-1}Z(j)}Z(k)\rangle,
\label{eq:string_order_parameter}
\end{equation}
where $r=|i-k|$ and $Z(i)=1-2a^\dagger_i a_i$. We consider $i=2$ and $k=L-1$ to avoid the edge sites. We calculate $O_{\text{str}}$ for a system with $L=200$ sites and plot in Fig.~\ref{fig:invariant_correlation_matrix}(a) denoted as red stars. It is clear that before $V=0$, the string order parameter is finite and vanishes after $V=0$ which agrees well with the topological invariant (blue diamonds). However, the string order parameter remains finite in the CDW-II phases~\cite{ebh_hidden_order,rossini_ebh,Batrouni_ebh} with opposite signs in the two regimes of interaction $V$. The change of sign  is a consequence of two differing spatial arrangements of particles in the CDW-II phases for $V>0$ (i.e., $01100110$) and $V<0$ (i.e., $11001100$). 

We also complement our findings by computing the density-density correlation function which is an experimentally relevant quantity~\cite{browyes,Mondal_topology}. This is
given by 
\begin{eqnarray}
 \Gamma_{i,j} ~=~ \langle Z_iZ_j\rangle ~-~ \langle Z_i \rangle 
 \langle Z_j \rangle.
 \label{eq:correlation_matrix}
\end{eqnarray}
We plot $\Gamma_{i,j}$ for $V=-1.0$ and $V=1.0$ in Figs.~\ref{fig:invariant_correlation_matrix} (c) and (d) respectively. The presence of finite correlation matrix elements at the two corners, i.e., the two isolated blue regions at opposite ends of the correlation matrix shown in Fig.~\ref{fig:invariant_correlation_matrix} (c) indicates the existence of edge states for $V=-1.0$ and hence the topological nature of the BO$_\pi$ phase. In contrast, Fig.~\ref{fig:invariant_correlation_matrix} (d) shows no such edge states, indicating the trivial nature of the BO$_0$ phase for $V=1.0$. 

\begin{figure}[t]
\centering
\includegraphics[width=0.8\linewidth]{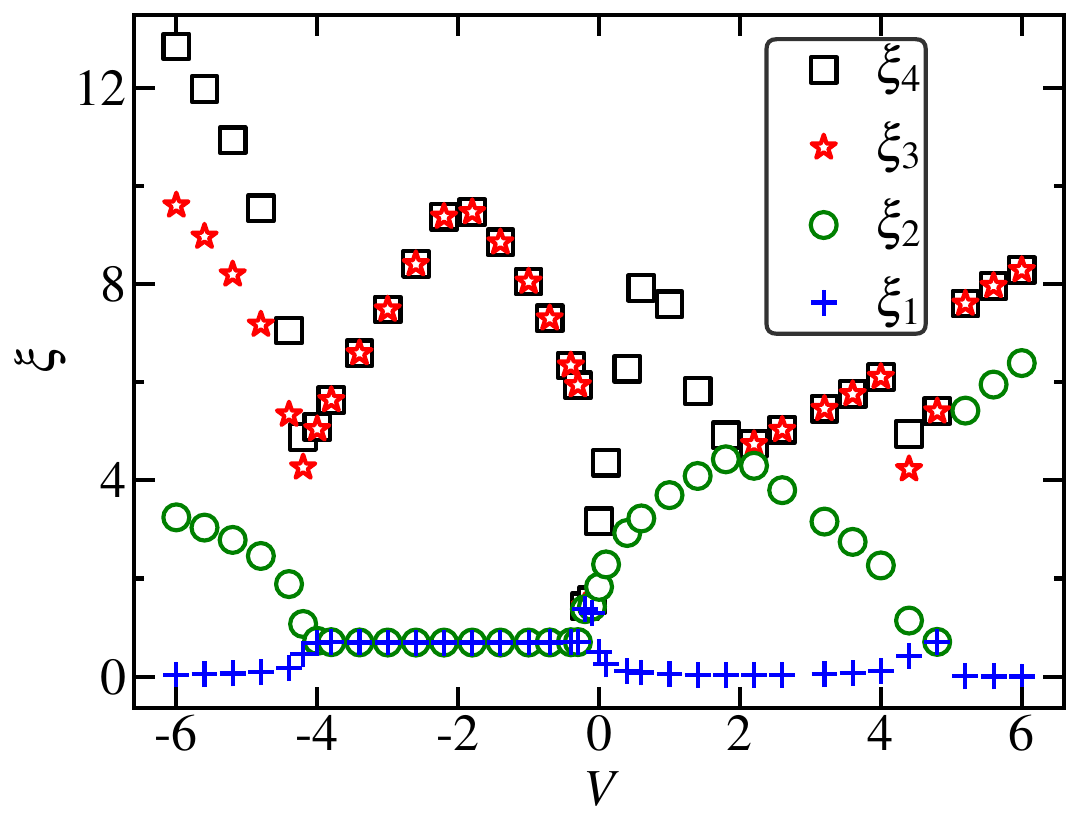}
\caption{The lowest four values of the entanglement spectrum ($\xi_1,~\xi_2,~\xi_3,~\xi_4$) as a function of $V$, for $L=200$.}
\label{fig:ent_spectrum}
\end{figure}

In addition to the above diagonistics, the degeneracy in the entanglement spectrum of the ground state serves as an excellent tool for identifying topological order in a system as it is related to the number of edge excitations in the system~\cite{haldane,kawakami,pollman_entropy,sjia}. The entanglement spectrum is obtained by logarithmically rescaling the eigenvalues of the reduced density matrix corresponding to a specific bi-partition of the system. For example, if $l$ is the length of the left block in the DMRG calculation, the reduced density matrix is given by $\rho_l = \text{Tr}_{L-l}|\psi\rangle\langle\psi|$, where $|\psi\rangle$ is the ground state wave function. In our case, the singular values at the $l^{th}$ bond in the MPS wave function of the ground state is related to the eigenvalues of the reduced density matrix as $\lambda_i = \Lambda_i^2$, where the $\lambda_i$'s are the eigenvalues of the reduced density matrix and the $\Lambda_i$'s are the singular values in the MPS wave function. With this idea we calculate the entanglement spectrum as
\begin{equation}
 \xi_i ~=~ - ~\text{ln}(\Lambda_i^2),
\label{eq:entanglement_spectrum}
\end{equation}
where the $\Lambda_i$'s correspond to the singular values at the central bond of the system in the MPS wave function of the ground state. We plot the lowest four values of the entanglement spectrum $\xi$ as a function of $V$ in Fig.~\ref{fig:ent_spectrum} for a system size of $L=200$. We find a two-fold degeneracy in the region $-4.3\lesssim V< 0$ apart from which the degeneracy is lifted for all the values of $V$. This indicates the topological nature of the BO ground state in the region $-4.3\lesssim V<0$. This diagnostic also complements the non-topological nature of the CDW-II phase arising for the $V\lesssim -4.3$ regime, claimed earlier.

The above analysis concretely establishes a topological phase transition from the BO$_\pi$ to BO$_0$ phase as the nature of the NN interactions is altered. In the following, we discuss the signatures of this topological phase transition in the context of Thouless charge pumping that is accessible in existing quantum simulators. 

\subsubsection*{Thouless Charge Pumping}
\begin{figure}[b]
\centering
\includegraphics[width=0.8\linewidth]{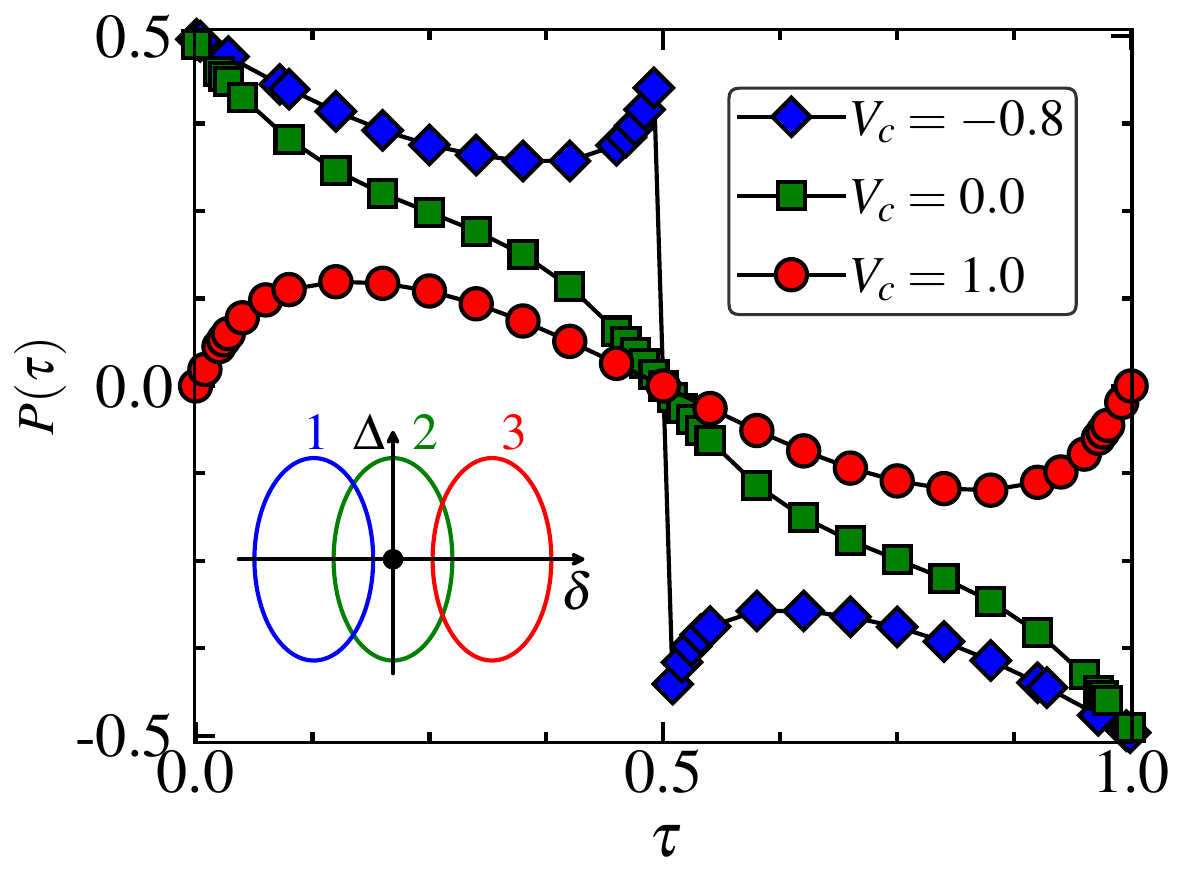}
\caption{Evolution of the polarization $P(\tau)$ as a function of $\tau$ for a system with $L=200$ sites, shown for three distinct pumping cycles (1, 2, 3) corresponding to different values of $V_c$ ($=-0.8,~0.0,~1.0$) as indicated in the inset. In all three cases, we set $A_\delta=0.6 ~\text{and}~ A_\Delta=0.6$. Blue diamonds, green squares, and red circles represent the polarization changes for cycles 1, 2, and 3, respectively. Robust charge pumping occurs in cycle 2, as it encloses the topological phase transition point $V=0.0$. In contrast, pumping breaks down (since there is a discontinuity in the polarization at $\tau = 0.5$) in cycle 1, whereas no charge is pumped during cycle 3.}
\label{fig:polarization}
\end{figure}
Here we propose a quantity that dynamically probes and characterizes the aforementioned topological phase called Thouless charge pumping or topological
charge pumping (TCP)~\cite{Thouless1983}. TCP reveals a profound connection between topology and quantum transport, describing the quantized transport of particles through a cyclic and periodic modulation of system parameters without any external bias. The system parameters include one that drives the topological phase transition and an additional inversion symmetry-breaking term, ensuring that the bulk gap remains open throughout the pumping cycle — a necessary condition for particle pumping. The pumping cycle involves the adiabatic evolution of the system from a topological phase to a trivial phase and subsequently back to the initial topological phase again, with the cyclic path encircling the topological phase transition point. Due to the topological nature of the pump, an integer number of particles are transported across the system, and the number of particles transported is directly linked to the topological invariant of the system. Though originally proposed in the context of non-interacting systems, TCP has now been studied in systems of interacting particles. Several theoretical and experimental studies have adopted TCP as one of the crucial tools for revealing the signature of topological phases in both non-interacting as well as interacting systems~\cite{Lohse2016,Takahashi2016pumping,monika_review,Hayward2018,rice_mele,monika_review,Asboth2016_rm,bound_pump1,bound_pump3,juliafare_quantum,Kuno2017,bertok_pump,mondal_phonon,Hayward2018,spin_pumping,pumping_quasicrystals,Taddia2017,pumping_1d,hubbarad_thouless_pump,qubit_pumping,padhan_ladder,rajashri_ladder,seba_nphy_pumming,seba_nature_pumping,pumping_esslinger,pump_reversal_esslinger}. In one dimension, the pumping protocol is given by the celebrated Rice-Mele model. For our system, the corresponding Rice-Mele model is written by adding a symmetry-breaking term, $\Delta$, defining our pumping Hamiltonian, which can now be expressed as
\begin{eqnarray}
 H(\tau)&=&- ~t\sum\limits_{i}({a}^\dagger_{i}{a}_{i+1}+\text{H.c.}) \nn \\
 && + ~(V_c-\delta(\tau)\sum\limits_{i}(-1)^{i+1}(n_i-\frac{1}{2})(n_{i+1}-\frac{1}{2}) \nn \\
 && +~ \Delta(\tau)\sum\limits_i(-1)^{i+1}n_i,
\label{eq:pumping_ham}
\end{eqnarray}
where $\tau$ is the pumping parameter. The parameter $\Delta$ ensures that the bulk gap remains open throughout the pumping cycle. $V_c$ is the origin of the pumping cycle in the $\delta-\Delta$ plane. $\delta(\tau)=A_\delta \text{cos}(2\pi \tau)$ and $\Delta(\tau)=A_\Delta\text{sin}(2\pi \tau)$ periodically modulate the NN interaction and the staggered potential term respectively. We consider three pumping cycles corresponding to three different values of $V_c$ as shown in the inset of Fig.~\ref{fig:polarization}. The parameter sets ($V_c,A_\delta,A_\Delta$) for these three cycles are taken to be ($-0.8t,0.6t,0.6t$), ($0.0t,0.6t,0.6t$) and ($1.0t,0.6t,0.6t$) for cycle 1,2, and 3 respectively. One should expect that pumping can only happen for cycle $2$ (green continuous line) as it winds around the topological phase transition point $V=0$. To study the TCP, we calculate a quantity called polarization given by
\begin{eqnarray}
 P(\tau) ~=~ \frac{1}{L} ~\sum\limits_{i=1}^{L} ~\langle\psi(\tau)|(i-i_0)n_i|\psi(\tau)\rangle,
 \label{eq:polarization}
\end{eqnarray}
where $i_0=(L+1)/2$ for a system of $L=200$ sites for all the three cycles. Here $\psi(\tau)$ represents the ground state wave function of the Hamiltonian defined in Eq.~(\ref{eq:pumping_ham}). The amount of charge pumped ($Q$) in a cycle is given as 
\begin{equation}
 Q = \int_0^{1} d\tau ~\partial_\tau P(\tau).
\label{eq:total_charge}
\end{equation}
Fig.~\ref{fig:polarization} clearly demonstrates a smooth change in the polarization from $0.5$ to $-0.5$, indicating a total charge transfer of $|Q|=1$ for cycle 2. This signifies robust and quantized pumping occurring in cycle 2 as it encloses the gap-closing topological phase transition point as mentioned earlier. In contrast, cycle 1 exhibits a breakdown of pumping (due to the appearance of a
discontinuity in the polarization), while cycle 3 shows no pumping at all~\cite{Hayward2018}. 
This analysis complements the topological phase transition between the two BO phases shown in the phase diagram [Fig.~\ref{fig:phasedia}(a)].

\subsection{Away from half-filling}

\begin{figure}[t]
\centering
\includegraphics[width=0.8\linewidth]{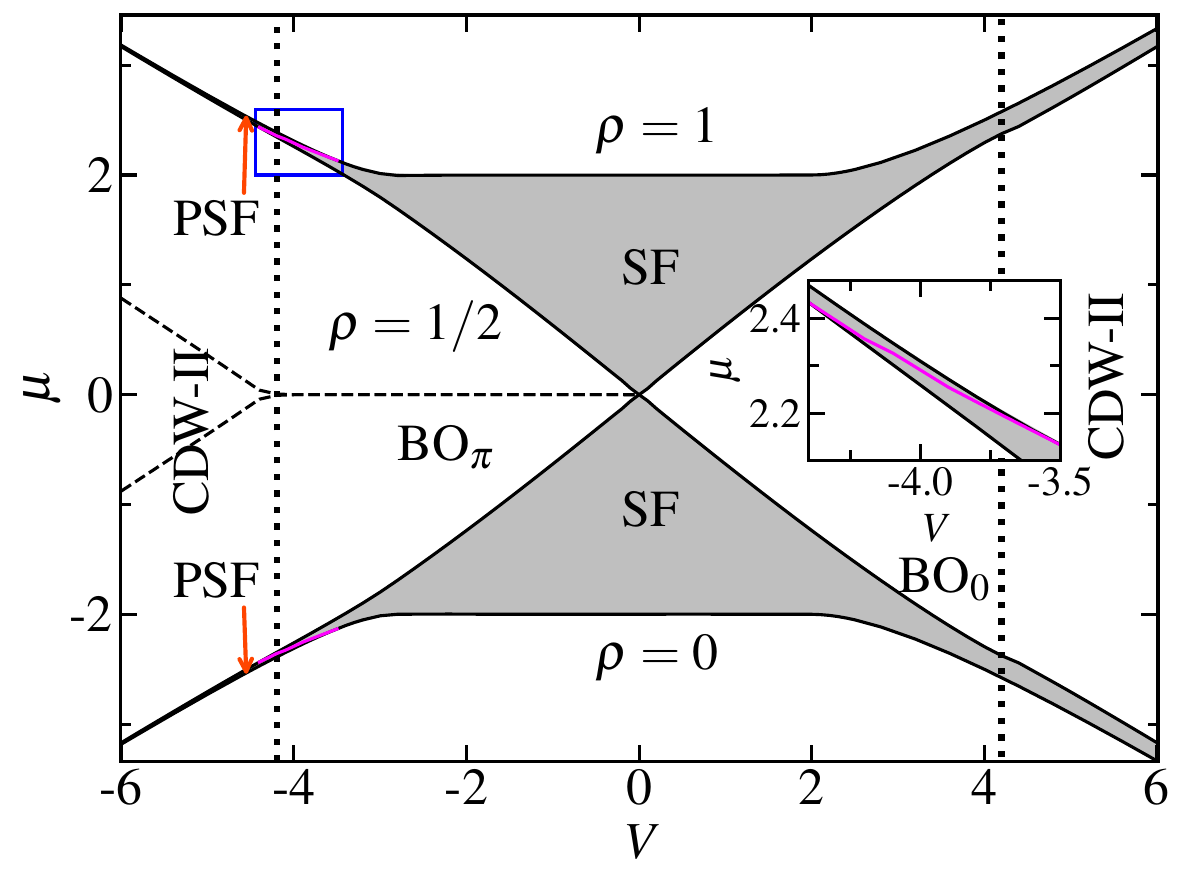}
\caption{The phase diagram of the model in $V-\mu$ plane by taking $V_1=-V_2=V$ under OBC. Dashed lines indicate the presence of edge states. The white regions in the middle enclosed by gray regions correspond to gapped phases at half-filling, while the white regions at $\rho=0$ and $\rho=1$ represent the empty and fully occupied states, respectively. Gray-shaded regions denote gapless phases, which are further classified based on excitation type: regions with single-particle excitations are identified as the SF phase, while regions with two-particle excitations correspond to the PSF phase. The solid magenta lines mark the boundaries between the SF and PSF phases. To clearly show this, we plot the zoomed-in portion demarcated by the blue rectangular box as shown in the inset.}
\label{fig:phase_diagram}
\end{figure}

\begin{figure}[t]
\centering
\includegraphics[width=1.0\linewidth]{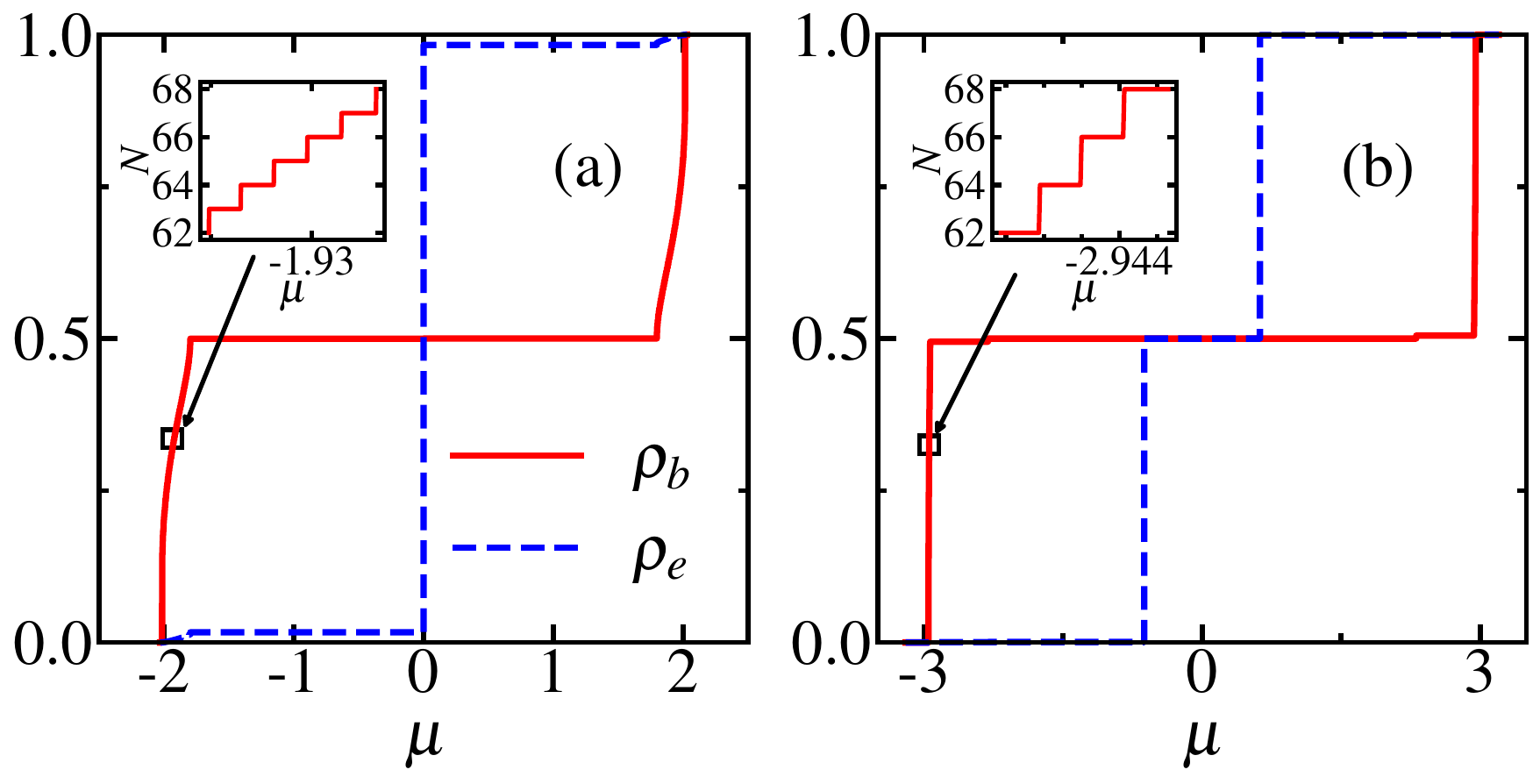}
\caption{The figure depicts the bulk and edge densities represented by $\rho_b$ (red solid lines) and $\rho_e$ (blue dashed lines) respectively as a function of $\mu$, for two different cuts through the phase diagram shown in Fig.~\ref{fig:phase_diagram}: (a) $V=-3.0$ and (b) $V=-5.5$ for a system size of $L=200$. This shows the nature of bulk phases (gapped or gapless) when the edge states are being filled in different parameter regimes. The inset shows the number of particles filled ($N$) as a function of chemical potential ($\mu$). The single-particle and two-particle jumps in the insets of (a) and (b) represent the existence of SF and PSF phases respectively.} \label{fig:rho_mu}
\end{figure}

After studying the ground state phase diagram in both the regimes of interactions at half-filling, we now extend our analysis to explore the system across all densities. For this purpose, we again focus on the same cut through the phase diagram that corresponds to the $V_1=-V_2=V$ line (gray dashed line in Fig.~\ref{fig:phasedia} (a)) as mentioned earlier. 
The phase diagram of the system in the $V-\mu$ plane is depicted in Fig.~\ref{fig:phase_diagram}. The gapped phases at $\rho=1/2$, namely the BO and the CDW-II phases, appear in the regions with $V>0$ and $V < 0$ and are separated by a vertical dotted line. At $V=0$ the gap vanishes and it corresponds to the topological phase transition point between the BO$_\pi$ and BO$_0$ phases at half-filling as discussed earlier. The gray regions on the upper and lower sides of the gapped phases are the gapless phases which appear for densities away from half-filling. The upper and lower white regions are the completely filled and completely empty (vacuum) states with $\rho=1$ and $\rho=0$ respectively. The boundaries of these phases are determined by following the analysis performed in Fig.~\ref{fig:rho_mu_gap}.
In this case however, we plot the bulk and the edge densities defined as 
\begin{equation}
 \rho_b=\frac{1}{L-2}\sum\limits_{i=2}^{L-1}\langle n_{i}\rangle,
\end{equation}
and
\begin{equation}
\rho_e= \frac{1}{2}[\langle n_{1}\rangle + \langle n_{L}\rangle],
\end{equation}
in Figs.~\ref{fig:rho_mu} (a) and (b) for two exemplary values of $V=-3.0$ and $V=-5.5$ as red solid lines and blue dashed lines, respectively.

The large plateaus in $\rho_b$ (red solid lines) in both the figures at $\rho=1/2$ is due to the gapped phases, and the shoulders across the plateaus correspond to the gapless phases. The length of the plateau at $\rho=1/2$ determines the gap at this filling, and the end points of the plateau determine the boundaries of the gapped phase appearing at $\rho=1/2$ corresponding to the particular cut in the phase diagram shown in Fig.~\ref{fig:phase_diagram}. Interestingly, there are two degenerate mid-gap zero energy modes appearing in the phase diagram starting from $V=0$ to $V\sim -4.3$ surpassing which these states become energetic, and these are indicated by the dashed lines at the center of the phase diagram. These are the signatures of edge states which have already been discussed in the previous section; 
they appear here as a consequence of the bulk topology. The degeneracy persists in the topological BO phase which then becomes non-degenerate because of the transition into the trivial CDW-II phase. 
This signature of edge states
can be captured from the $\rho-\mu$ plot shown in Fig.~\ref{fig:rho_mu}. The jump in $\rho_e$ by 1 when the bulk is gapped in Fig.~\ref{fig:rho_mu} (a) at $\mu=0$ is an indication of the presence of two zero energy edge states in this parameter regime. Similarly, the jump in $\rho_e$ by 0.5 in Fig.~\ref{fig:rho_mu} (b) while the bulk is gapped at finite $\mu$ represents the presence of finite-energy edge states.

We now examine the phases away from commensurate densities. As evident from the phase diagram [Fig.~\ref{fig:phase_diagram}] as well as from the $\rho-\mu$ curve [Fig.~\ref{fig:rho_mu}], the regions away from half-filling correspond to gapless phases. In some regions, the excitations are of single-particle type representing an SF phase, whereas in some other regions, the excitations are pair excitations representing a pair superfluid (PSF) phase. The PSF phase arises due to strong attractive interactions that favor the formation of bound nearest-neighbor pairs. To distinguish between the SF and PSF phases, we plot the particle number $N$ as a function of the chemical potential $\mu$ in the inset of Fig.~\ref{fig:rho_mu}, focusing on the region marked by the black rectangle for a system size of $L=200$ sites. The single-particle jump in $N$ with increase in $\mu$ for $V=-3.0$ (inset in Fig.~\ref{fig:rho_mu} (a)) shows the presence of an SF phase, whereas the two-particle jump for $V=-5.5$ (inset in Fig.~\ref{fig:rho_mu} (b)) shows the existence of the PSF phase. These features can also be understood from analytical arguments discussed in Appendix~\ref{app:appen_fullyoccupied}. The jumps in $\rho_b$ in steps of $1$ occurring when $\mu$ is slightly more
than $-2.0$ (Fig.~\ref{fig:rho_mu}(a)) is due to single particles appearing in the bulk when $\mu$ increases 
above $-2.0$ as discussed in Appendix~\ref{app:appen_fullyoccupied} (see Eq.~\eqref{bound2}).
The jumps in $\rho_b$ in steps of $2$ occurring when $\mu$ increases 
above $-2.94$ (Fig.~\ref{fig:rho_mu}(b)) is due to the appearance of two-particle bound states in the bulk. 
If we substitute $V=-5.5$ in Eq.~\eqref{bound5}, it gives a 
threshold equal to about $-2.95$ which matches well with the numerical results.

\begin{figure}[t]
\centering
\includegraphics[width=0.8\linewidth]{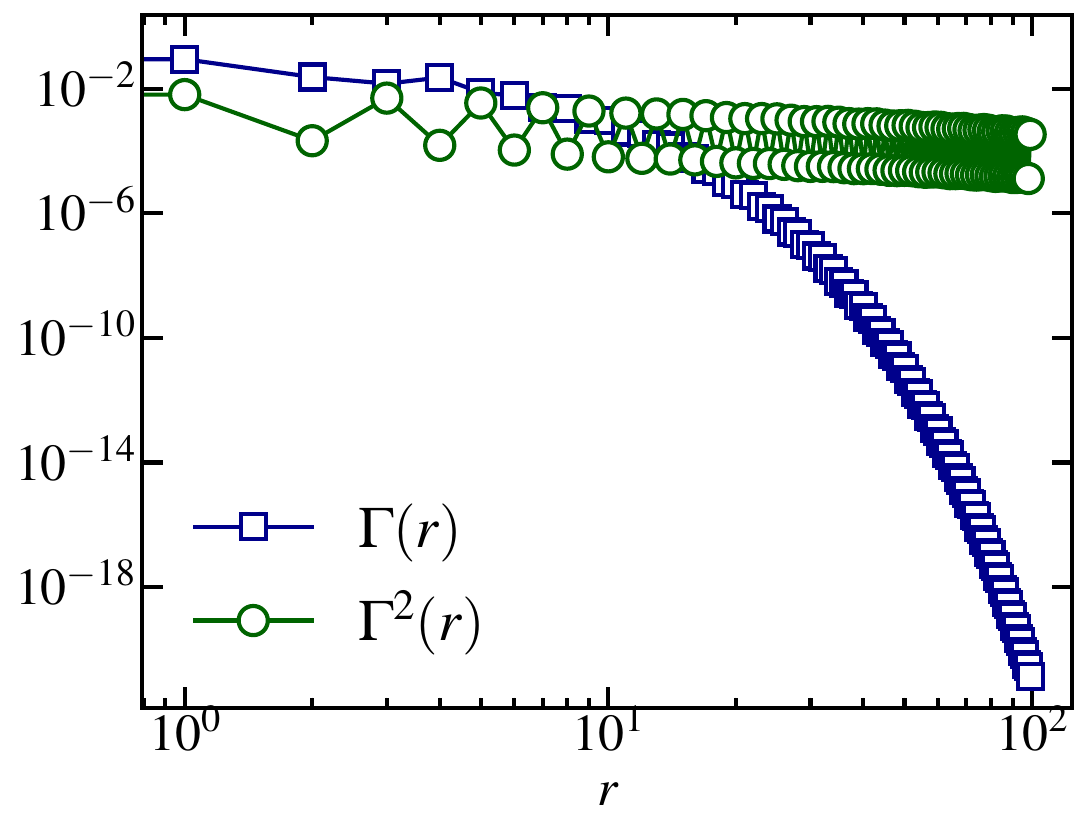}
\caption{The SF correlation function $\Gamma\text(r)$ and the PSF correlation function $\Gamma^\text{2}(r)$ plotted as a function of the distance ($r=|i-j|$) on a log-log scale at $\rho=0.25$ for $V=5.5$ for a system with size $L=200$. To avoid the boundary effects we only consider the lattice sites from $L/4$ to $3L/4$.}
\label{fig:correlation}
\end{figure}

The SF and PSF boundaries are separated by the magenta lines in the phase diagram shown in Fig.~\ref{fig:phase_diagram}.  Note that a similar SF-PSF transition line is observed for the $V > 0$ case; however, it is not seen from the $\rho$-$\mu$ curve obtained with OBC because of the choice of the boundary condition and the repulsive interaction in the first and the last bonds. More details can be found in Appendix~\ref{app:appen_fullyoccupied}. Nevertheless, the PSF nature can still be verified by examining the behavior of the PSF correlation function which is given by
\begin{equation}
 \Gamma^{2}(r=|i-j|) = \langle a_i^\dagger a_{i+1}^\dagger a_j a_{j+1}\rangle.
 \label{eq:psf_corr}
\end{equation}
The algebraic decay of the PSF correlation function $\Gamma^2(r)$ (green circles) in Fig.~\ref{fig:correlation} for $V=5.5$ at filling $\rho=0.25$ shows the existence of a PSF phase in this parameter regime. For comparison, we also plot the SF correlation function $\Gamma(r)$ (blue squares) defined in Eq.~\ref{eq:sf_corr} which shows an exponential decay confirming the existence of the PSF order in the system. 

\begin{figure}[b]
\centering
\includegraphics[width=0.8\linewidth]{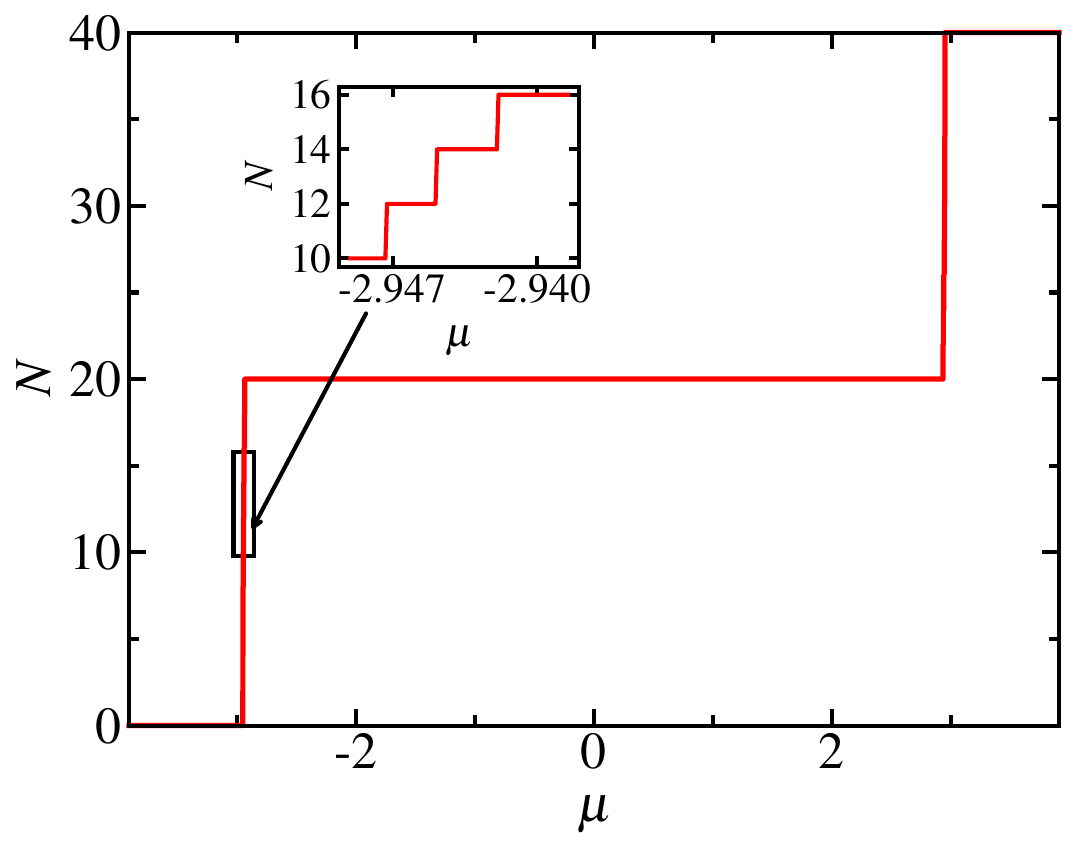}
\caption{The figure shows the number of particles filled ($N$) as a function of the chemical potential ($\mu$) for $V=5.5$ under PBC for a system size of $L=40$. The inset shows the zoomed-in region of the red curve which is enclosed within the black rectangle. The two-particle jump is clear from the inset which is a signature of the PSF phase.}
\label{fig:rho_mu_pbc}
\end{figure}

To address this issue arising because of OBC, the boundary condition can be changed to PBC which allows for a clearer examination of the $\rho-\mu$ curve. Hence, we analyze the $\rho-\mu$ curve at $V=5.5$ by employing PBC for a system size of $L=40$ and show the result in Fig.~\ref{fig:rho_mu_pbc}. The presence of a two-particle jump away from half-filling under PBC confirms the existence of the PSF phase in the bulk. 
The jumps occur due to the appearance of two-particle bound states in the bulk 
as discussed in Appendix~\ref{app:appen_fullyoccupied} (see Eq.~\eqref{bound5}).
In contrast, under OBC, a single-particle density jump is observed even within the PSF region. Owing to this specific artifact introduced by the boundary condition, the phase diagram in Fig.~\ref{fig:phase_diagram} is not symmetric about $V=0$ although the bulk physics remains the same on the two sides. We provide a qualitative understanding of this picture in Appendix~\ref{app:appen_fullyoccupied}; see the discussion around 
Eqs.~\eqref{bound6} and \eqref{bound7} of hole
or particle end modes which appear for $V>2.0$.

\begin{figure}[t]
\centering
\includegraphics[width=0.8\linewidth]{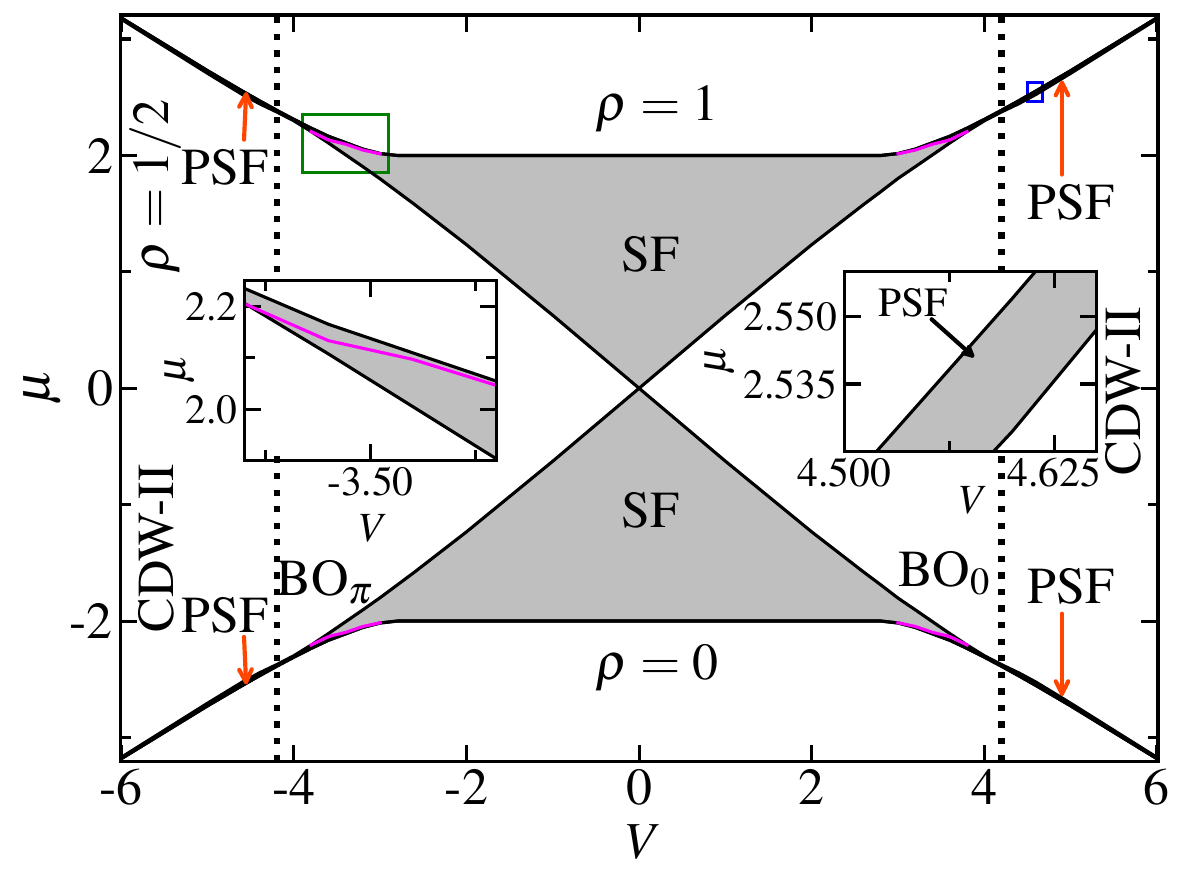}
\caption{Phase diagram of the model in the $V-\mu$ plane by taking $V_1=-V_2=V$ with PBC. All the phase boundaries are extrapolated by taking a maximum system size of $L=60$. All the regions have the usual meaning as mentioned in the phase diagram shown in Fig.~\ref{fig:phase_diagram}. The solid magenta lines mark the boundary between the SF and the PSF phases. To clearly show this, we plot the zoomed-in portion demarcated by the green rectangular box in the left inset. The rectangular region marked in blue in the upper right part of the diagram is enlarged in the right inset to show that the $\rho=1$ and $\rho=1/2$ regions are separated by a finite gray region that carries the PSF order. }
\label{fig:phase_diagram_pbc}
\end{figure}

To resolve this issue numerically, 
we once again compute the bulk phase diagram by implementing the PBC; this is presented in Fig.~\ref{fig:phase_diagram_pbc}. The maximum system size in this case has been taken to be $L=60$, and all the phase boundaries are extrapolated to the thermodynamic limit. The mid-gap edge modes are now absent in the $V<0$ regime, as the implementation of PBC eliminates the edge states. Now the phase diagram exhibits complete reflection symmetry about $V=0$. Additionally, there are four magenta lines in the phase diagram that mark the boundaries between the SF and PSF phases. One can clearly see that at higher interaction strengths, the boundary region between the vacuum state ($\rho=0$) and the half-filled state ($\rho=1/2$) (or between the fully occupied state ($\rho=1$) and the half-filled state) appears narrow and seemingly merge to a single line. However, a finite intermediate gray region remains, consisting of two-particle excitations, which defines the PSF phase as illustrated in the right inset of Fig.~\ref{fig:phase_diagram_pbc}.

In Appendix~\ref{app:appen_fullyoccupied}, we present an understanding of the phase boundary 
below the fully filled state with $\rho=1$ in 
Fig.~\ref{fig:phase_diagram_pbc}. We show that the boundary is given by
$\mu = 2t$ if $|V| \le 2 \sqrt{2} t$ which results from one-hole states,
and by
\beq \mu ~=~ t ~\left[ \sqrt{ \frac{V^2}{4t^2} ~-~ 1} ~+~ \frac{1}{
\sqrt{ \frac{V^2}{4t^2} ~-~ 1}} \right], \eeq
if $|V| > 2 \sqrt{2} t$ which results from two-hole bound states and these boundaries agree well with the numerical calculations (Fig.~\ref{fig:phase_diagram_pbc}).
Similar expressions hold for the phase boundary above the empty state with $\rho=0$ in Fig.~\ref{fig:phase_diagram_pbc} (see Eq.~\eqref{bound5}).

 The above analysis provides a comprehensive study of the phases emerging at all densities. Apart from the fully empty and fully occupied states, a gapped phase appears at half-filling, as previously discussed. At all other densities, the system exhibits gapless nature. In certain regions, single-particle excitations dominate, characterizing the SF phase, while other regions are dominated by bound pair excitations and hence called the PSF phase. In the following, we discuss how such a model can be realized in experiments. 

\subsection{Experimental scheme}

\begin{figure}[t]
\centering
\includegraphics[width=0.8\linewidth]{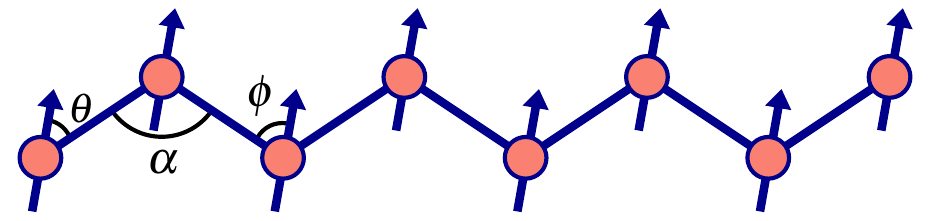}
\caption{Array of dipoles representing the model under consideration. The angles $\theta$ and $\phi$ control the strength and signs of the interactions $V_1$ and $V_2$, respectively. $\alpha$ decides the strength of the next-nearest-neighbour hopping strengths. }
\label{fig:cartoon_v_-v}
\end{figure}

\begin{figure}[b]
\centering
\includegraphics[width=0.8\linewidth]{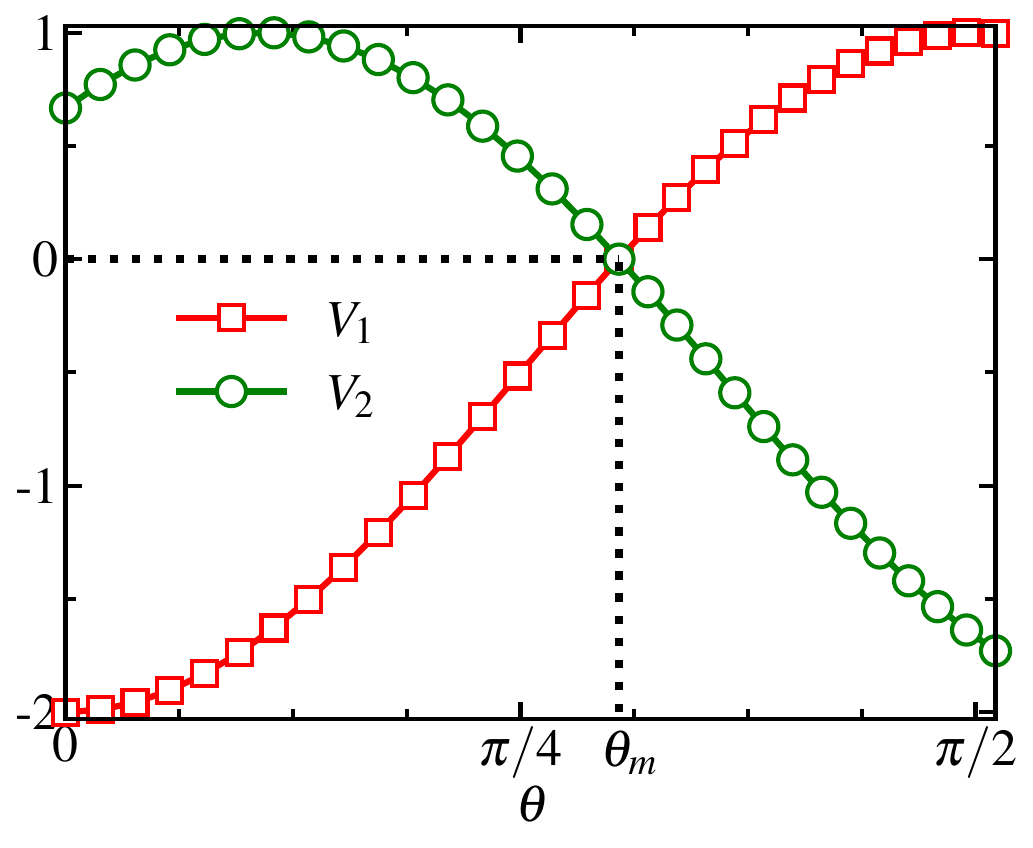}
\caption{Tunable NN interactions $V_1$ (red squares) and $V_2$ (green circles) are plotted as a function of $\theta$. Here, the values of the interactions are scaled according to the prefactor $\frac{q^2}{4\pi\epsilon_0r^3}$ given in Eq.~\ref{eq:dipolar_interaction}.}
\label{fig:varying_theta}
\end{figure}
In this subsection, we briefly discuss the possible experimental realization of the model with tunable NN interactions. As already mentioned in the introduction, the NN interactions have been successfully realized in various quantum simulators. Here, we show that a properly arranged array of dipolar atoms loaded in an optical lattice can be a suitable platform to achieve the tunability of the potentials $V_1$ and $V_2$ that is required to observe the topological phase transition. In Fig.~\ref{fig:cartoon_v_-v}, we show that the atoms arranged in the zig-zag pattern can selectively allow the atoms to hop between the NN sites while the hopping to the next-nearest-neighbor sites can be avoided by increasing the angle $\alpha$. We assume that the dipoles can be oriented in the $XY$-plane. The dipole-dipole interaction between the two dipoles is given by
\begin{eqnarray}
 V_{dd}=\frac{q^2}{4\pi\epsilon_0 {r}^3}(1-3\text{cos}^2(\Phi)),
\end{eqnarray}
where $\Phi$ is the angle made by the dipoles with the line joining the two dipoles~\cite{pfau}. Consequently, the NN interaction strengths are given by 
\begin{eqnarray}
 V_{1}=\frac{q^2}{4\pi\epsilon_0 {r}^3}(1-3\text{cos}^2(\theta)), \nn\\~~~\text{and} ~~~V_{2}=\frac{q^2}{4\pi\epsilon_0 {r}^3}(1-3\text{cos}^2(\phi)),
 \label{eq:dipolar_interaction}
\end{eqnarray}
where $\theta$ and $\phi$ are the angles specified in Fig.~\ref{fig:cartoon_v_-v}. Hence to tune the interaction between the NN dipoles as required by the model under consideration, $V_1$ and $V_2$ can be tuned by uniformly rotating all the dipoles. Here, we set $\alpha=2\theta_m$, where $\theta_m$ is the "magic angle" — the angle at which the dipole-dipole interaction strength vanishes which is given by $\text{cos}^{-1}(\frac{1}{\sqrt{3}})$. In this configuration, the angle 
$\phi$ is given by $2\theta_m-\theta$. By externally controlling the dipole orientation in the $XY$-plane, it is possible to adjust $\theta$, which in turn modifies $\phi$. It then becomes possible to simultaneously set both the angles $\theta$ and $\phi$ to be $\theta_m$ which makes both the NN interaction strengths zero. Fig.~\ref{fig:varying_theta} presents the variation of $V_1$ and $V_2$ as a function of $\theta$, assuming the prefactor in Eq.~\ref{eq:dipolar_interaction} to be unity. Initially, $V_1$ (red squares) is negative, while $V_2$ (green circles) is positive. As $\theta$ reaches $\theta_m$, both interactions vanish. Further increasing $\theta$ beyond this point reverses their signs. This controlled variation of interaction strengths enables the realization of topological phase transitions in the system.

\begin{figure}[t]
\centering
\includegraphics[width=0.8\linewidth]{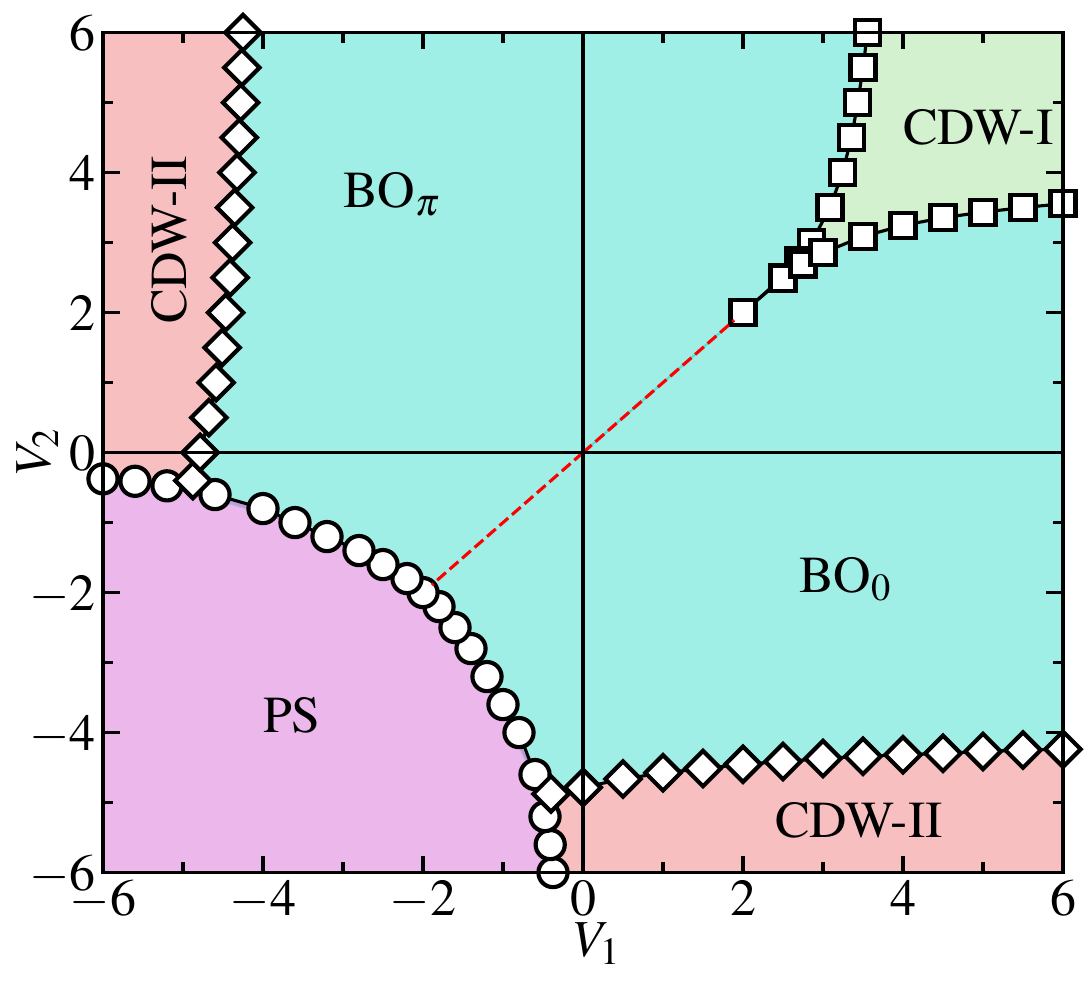}
\caption{Phase diagram in the full $V_1$–$V_2$ parameter space at half-filling, showing five distinct phases: bond order (BO), superfluid (SF), charge-density wave-I (CDW-I), charge-density wave-II (CDW-II), and phase separation (PS). White squares mark the BO to CDW-I phase boundary, white diamonds indicate the BO to CDW-II boundary, and the region enclosed by white circles corresponds to the PS phase. The red dashed line marks the SF phase.} 
\label{fig:fullphasedia}
\end{figure}

\begin{figure*}[t]
\centering
\includegraphics[width=0.8\linewidth]{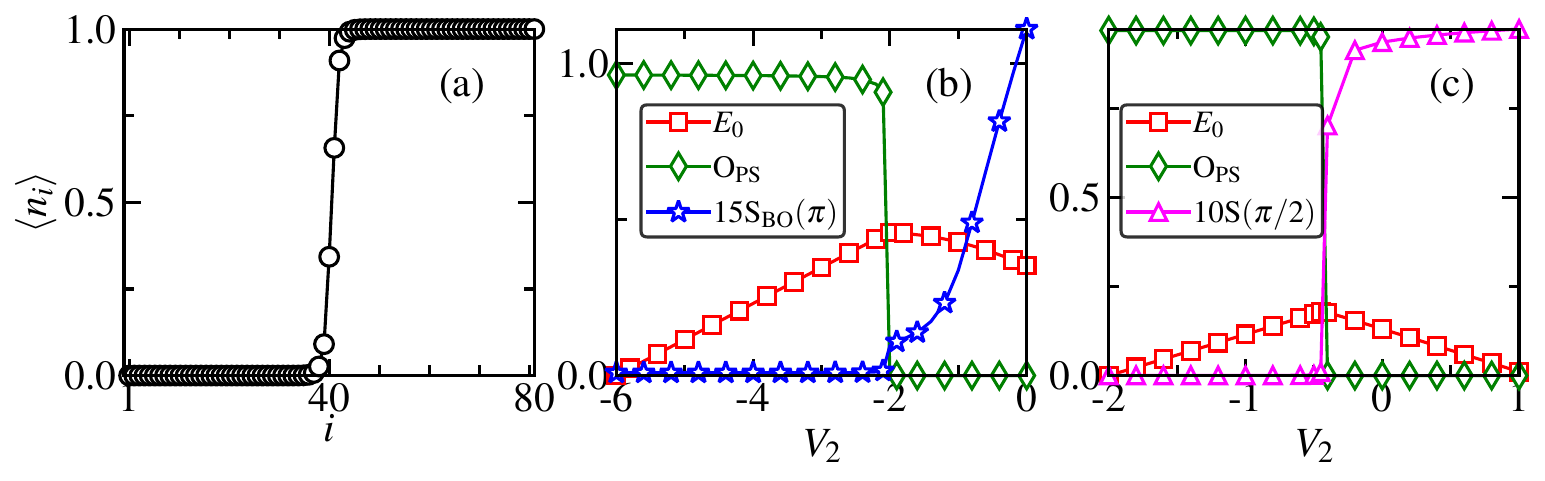}
\caption{a) On-site particle density $\langle n_i \rangle$ vs. site index $i$ for $V_1 = -2.0$ and $V_2 = -4.0$, calculated on a system of
$L=80$ with OBC. (b) compares O$_\text{PS}$ (green diamonds) and S$_{\text{BO}}(\pi)$ (blue stars) plotted as a function of $V_2$, for $V_1=-2.0$. (c) compares O$_\text{PS}$ (green diamonds) and S$(\pi/2)$ (magenta triangles) plotted as a function of $V_2$ for $V_1=-5.4$.  In order to avoid boundary effects arising due to OBC, we consider the sites from $L/10$ to $9L/10$ for the calculation of O$_\text{PS}$.  For clarity, S$_{\text{BO}}(\pi)$ and S$(\pi/2)$ are scaled by factors of $15$ and $10$, respectively. The ground-state energy density $E_0$ (red squares) is plotted as a function of $V_2$ in (b) and (c) for $V_1=-2.0$ and $-5.4$, respectively. The values of $E_0$ have been plotted with vertical offsets of $0.95$ and $0.91$ added in panels (b) and (c), respectively, to enhance clarity.} 
\label{fig:1qpt}
\end{figure*}

\section{Complete Phase Diagram at half-filling}
\label{fullphase}

In this section, we present the full phase diagram in the $V_1-V_2$ parameter space in Fig.~\ref{fig:fullphasedia}, to provide a comprehensive overview of the ground state phases of the model. The second ($V_1<0,V_2>0$) and the fourth ($V_1>0,V_2<0$) quadrants have been analyzed in detail in the preceding section, while the first quadrant ($V_1,V_2>0$) was previously studied in Ref.~\cite{Mondal_topology, harsh_xxz}. It has already been shown that the first quadrant contains both the BO$_0$ and BO$_\pi$ phases, separated by a topological phase transition in the weak interaction regime (indicated by the red dashed line in Fig.~\ref{fig:fullphasedia}). In addition to this, a charge-density wave phase (CDW-I), characterized by the occupation pattern of $\cdots 01010101 \cdots$, appears within the region enclosed by white squares with increase in interactions.

To understand the scenario in the third quadrant ($V_1,V_2<0$), we first examine the symmetric line which corresponds to $V_1=V_2$. Along this line, the model maps exactly to a spin-$1/2$ $XXZ$ chain, where $V_2 (=V_1)$ controls the strength of anisotropy and the $XX$ interaction strength is $-2t$. The $XXZ$ model, solvable via the Bethe Ansatz~\cite{Sutherland}, exhibits a gapless $XY$-ordered phase (equivalent to a SF phase in the bosonic language) for $-2t<V_2<2t$~\cite{Cabra}, as indicated by the red dashed line in Fig.~\ref{fig:fullphasedia}. 
At $V_2=-2t$, a first-order quantum phase transition is known to occur from the $XY$ ordered phase to a ferromagnetic phase, which is also the phase-separated (PS) phase in the bosonic context. This PS phase features a clustered density profile of the form $11110000$, and the phase transition from the SF to the PS phase is signaled by a kink in the ground state energy $E$, indicating a non-analyticity in its derivative~\cite{1qpt}. From our DMRG simulations, we explore this phase transition and the entire phase diagram in the $V_1-V_2$ plane corresponding to the third quadrant as depicted in Fig.~\ref{fig:fullphasedia}. We obtain that the BO phases of the second and the fourth quadrants extend to the third quadrant and are separated by the topological phase transition line at $V_1=V_2$ (red dashed line) up to $V_1=V_2=2$. The CDW-II phases also extend to this quadrant at smaller values of interactions. Eventually, all these phases undergo transitions to the PS phase at the phase boundary denoted by circles. In the following, we characterize these phases and the transitions between them.


While we characterize the BO and the CDW-II phases by their respective order parameters discussed above, we identify the PS phase from the spatial distribution of the onsite density $\langle n_i\rangle$. In Fig.~\ref{fig:1qpt}(a) we plot the $\langle n_i\rangle$ vs $i$ for $V_1=-2.0$ and $V_2=-4.0$ which clearly shows the accumulation of all the particles in a region of the lattice indicating a phase separation. To quantify the PS phase, we define an order parameter as
\begin{equation}
\text{O}_{\text{PS}} = \frac{1}{L/2} \left| \sum_{i = 1}^{L/2} \langle n_i \rangle - \sum_{i = L/2 + 1}^{L}\langle n_i \rangle \right|,
\label{eq:ps_op}
\end{equation}
which remains finite only within the PS phase and vanishes elsewhere.
In Fig.~\ref{fig:1qpt}(b) and (c), we plot O$_\text{PS}$ (green diamonds) as a function of $V_2$ for two representative cuts in the phase diagram [Fig.~\ref{fig:fullphasedia}] at $V_1=-2.0$ and $V_1=-5.4$, corresponding to BO to PS and CDW-II to PS transitions, respectively. To highlight the transitions to the BO and the CDW-II phases, we also plot S$_{\text{BO}}(\pi)$ (blue stars) and S$(\pi/2)$ (magenta triangles) along with O$_\text{PS}$. In both the figures, it is evident that O$_\text{PS}$ becomes finite near the point where the corresponding structure factors vanish, which clearly indicates the onset of the PS phase.

While the order parameters described above i.e. S$_{\text{BO}}(\pi)$, S$(\pi/2)$, and O$_\text{PS}$ are sufficient to describe different phases and the transitions between them, we determine the phase transition boundary to the PS phase (circles) by tracking the point ($V_2$) at which the slope in the ground state energy ($E$) changes its sign. To this end, we plot the ground state energy density $E_0=E/L$ (red squares) as a function of $V_2$ in Fig.~\ref{fig:1qpt}(b) and (c) for fixed values of $V_1=-2.0$ and $-5.4$, respectively. The resulting curves display a distinct kink, marking the phase transition points.

In the limit of large $|V_1|$ (or large $|V_2|$), the boundary between one of the CDW-II phases and the PS phase can be found using second-order perturbation theory; this is presented in Appendix~\ref{app:appen_degenpert2}. 
For instance, as $V_1 \to - \infty$, we find that there is a first-order phase transition
of a classical Ising model at a line
given by 
\beq V_2 ~=~ \frac{2t^2}{V_1}. \eeq
This asymptotically approaches $V_2 = 0$ from
the negative side as $V_1 \to - \infty$, as we see 
near the middle of the left edge of Fig.~\ref{fig:fullphasedia}.

\section{Conclusion}\label{conclusion}

We have numerically investigated the ground state properties of a one-dimensional lattice of hardcore bosons or spinless fermions with tunable NN interactions. While the non-interacting counterpart of the system is a trivial gapless phase, we have shown that by allowing attractive and repulsive interactions on the alternate NN bonds of the lattice, a topological phase can be established. We have also shown that by tuning the NN interaction appropriately, a topological to trivial phase transition occurs at a critical point where the interactions vanish. Apart from this, we have also obtained a symmetry broken phase for stronger interaction strengths. While we have revealed that the bulk of the topological and the trivial phases exhibit finite bond ordering, the symmetry broken phase exhibits a CDW order. We have obtained the phase diagram depicting these phases and the transitions between them at half-filling with tunable NN interactions. Furthermore, by extending our analysis beyond half-filling, we have explored the phase diagram at all densities and identified both single-particle and bound-pair excitations at incommensurate fillings, leading to the superfluid (SF) and pair-superfluid (PSF) phases. We have also provided analytical arguments for specific scenarios in support of the numerical results. Finally, we discussed the experimental feasibility of realizing the model under consideration in quantum simulation platforms, such as Rydberg atom and ultracold atomic set ups. To complete the study, we discussed the experimental feasibility of realizing the model under consideration in quantum simulation platforms, such as Rydberg atom and ultracold atomic setups. Finally, by systematically exploring all possible combinations of attractive and repulsive NN interactions, we presented the full interaction-driven phase diagram, offering a comprehensive understanding of the emergent quantum phases and their transitions.

\section{Acknowledgements}
We thank Immanuel Bloch, Rejish Nath, Subhro Bhattacharjee and Ashirbad Padhan for useful discussions. We also thank Biswajit Paul and Sanchayan Banerjee for useful comments on the manuscript. 
We acknowledge discussions in the ICTS
program titled ``A Hundred Years of Quantum Mechanics"
(code: ICTS/qm100-2025/01). T.M. acknowledges support 
from the Science and Engineering Research Board (SERB), Government of India, 
through project No. MTR/2022/000382 and No. STR/2022/000023. D.S. 
acknowledges support from SERB, Government of India, through 
project No. JBR/2020/000043.

\appendix

\section{Perturbation theory in the limit $V_2 \to \infty$}
\label{app:appen_degenpert1}

In this Appendix we will provide a derivation of Eq.~\eqref{v1v2} in the main
text. Given the Hamiltonian in Eq.~\eqref{eq:spin_ham}, we will assume that
$V_2 \gg t$ and $|V_1|$, and will go up to second order in perturbation theory.
We begin by considering only the terms with $V_2$ which constitutes the
unperturbed Hamiltonian. The ground state of
this part of the Hamiltonian in Eq.~\eqref{eq:spin_ham} has a large
degeneracy: for every {\it even} value of $i$, the pair of spins on sites $i$ 
and $i+1$ must be either in the state $| \upa_i \dna_{i+1} \ra$
or in the state $| \dna_i \upa_{i+1} \ra$. We therefore define 
a new lattice with sites labeled as $j=i/2$ and spin-1/2 operators ${\vec \tau}_j$ which sit at the bond 
$(i,i+1)$ of the original lattice such that the above two states correspond to $\tau_j^z = 1$ and
$\tau_j^z = -1$ respectively. 

In the space of degenerate ground states described above, we see that
the $V_1$ term in Eq.~\eqref{eq:spin_ham} effectively acts as $-(V_1/4) 
\tau_j^z \tau_{j+1}^z$. For instance, acting
on the state $| \tau_j^z = 1, \tau_{j+1}^z =1 \ra = | \upa_i 
\dna_{i+1} \upa_{i+2} \dna_{i+3} \ra$, the term $V_1 S_{i+1}^z 
S_{i+2}^z$ gives $(-V_1/4) | \tau_j^z = 1, \tau_{j+1}^z = 1 \ra$.

Next, the term in Eq.~\eqref{eq:spin_ham} of the form
\beq - ~2 t ~(S_i^x S_{i+1}^x ~+~ S_i^y S_{i+1}^y) ~=~ - ~t
~(\sigma_i^+ \sigma_{i+1}^- ~+~ \sigma_i^- \sigma_{i+1}^+) \eeq
flips the state $\tau_j^z = 1$ to $\tau_j^z = -1$, and vice versa.
Hence this term effectively acts as $-t ~\tau_j^x$.

However, the term in Eq.~\eqref{eq:spin_ham} of the form
\beq - ~2 t ~(S_{i-1}^x S_i^x ~+~ S_{i-1}^y S_i^y) ~=~ - ~t
~(\sigma_{i-1}^+ \sigma_i^- ~+~ \sigma_{i-1}^- \sigma_i^+), \label{ipm}
\eeq
where $i$ is even, does not have any matrix elements between states
in the space of degenerate ground states. But this term contributes 
to second order as follows. Acting on the state $| \tau_{j-1}^z = 1, 
\tau_j^z = 1 \ra = | \upa_{i-2} \dna_{i-1} \upa_i \dna_{i+1} \ra$,
the term in Eq.~\eqref{ipm} gives $-t | \upa_{i-2} \upa_{i-1} \dna_i 
\dna_{i+1} \ra$ which is an excited state whose energy is $V_2$ 
larger than the energy of the degenerate ground states. Another application 
of Eq.~\eqref{ipm} brings the excited state back to 
$| \tau_{j-1}^z = 1, \tau_j^z = 1 \ra$, with a coefficient
equal to $(-t)^2/(-V_2) = - t^2/V_2$ as given by second-order perturbation
theory; the energy denominator equal to $-V_2$ comes from 
$E_{ground} - E_{excited}$. A similar terms appears
for the state $| \tau_{j-1}^z = -1, \tau_j^z = -1 \ra$. However,
no such terms appear for the states $| \tau_{j-1}^z = 1, \tau_j^z = -1 
\ra$ and $| \tau_{j-1}^z = -1, \tau_j^z = 1 \ra$ since Eq.~\eqref{ipm}
gives zero on these states. Hence, Eq.~\eqref{ipm} gives rise to a
term in the effective Hamiltonian given by $(-t^2/(2V_2)) (\mathbb{I} ~+~
\tau_{j-1}^z \tau_j^z)$. 

Hence, ignoring some constants, the low-energy
effective Hamiltonian is given by
\beq H_{eff} ~=~ \sum_j ~[(- ~\frac{V_1}{4} - \frac{t^2}{2V_2})~
\tau_{j-1}^z \tau_j^z ~-~ t ~\tau_j^x]. \label{heff1} \eeq
This describes the transverse field Ising model
(called the quantum Ising model). It is well-known that
this model has a second-order phase transition when the coefficients of the
two terms in Eq.~\eqref{heff1} are equal in magnitude, i.e., when 
$t = |(V_1/4) + (t^2/(2V_2))|$. For $V_1 < 0$, this gives the line
\beq V_1 ~=~ - ~4t ~-~ \frac{2t^2}{V_2}. \eeq 
This is Eq.~\eqref{v1v2} in the main text.

\section{Boundaries close to $\rho = 1$ and $\rho =0$}
\label{app:appen_fullyoccupied}

In this Appendix, we will consider the Hamiltonian in Eq.~\eqref{eq:ham}
with $V_1 = - V_2 = V$, and add a chemical potential term $- \mu \sum_i n_i$. 
We will derive the forms of the boundaries close to
the completely filled and completely empty states
that we see in Fig.~\ref{fig:phase_diagram} for open boundary
conditions and Fig.~\ref{fig:phase_diagram_pbc} for periodic boundary conditions.

To understand Fig.~\ref{fig:phase_diagram_pbc}, we will consider an $L$-site system 
with periodic boundary conditions so that momentum eigenstates can be defined.
The fully occupied state in which the number of particles is $L$
is an eigenstate of Eq.~\eqref{eq:ham} with energy equal to $E_0 = -\mu L$.
Next, we consider a one-hole state in which the number of particles is $L-1$.
Defining a state with a hole located at site $j$ as $|j \ra$, we
find that 
\beq H | j \ra ~=~ - \mu (L-1) ~| j \ra ~-~ t ~| j-1 \ra ~-~ t | j+1 \ra.
\label{hj} \eeq
Next, we define a momentum eigenstate as
\beq | k \ra ~=~ \sum_i ~e^{ijk} ~| j \ra. \eeq
We then find that $H | k \ra = - (\mu L - \mu + 2 t \cos k) | k \ra$.
Subtracting the energy $E_0$ of the fully occupied state, we see that the
dispersion of a single hole is given by
\beq E_1 (k) ~=~ \mu ~-~ 2 t ~\cos k. \label{Ek} \eeq
This implies that the one-hole state has lower energy than the fully 
occupied state below the line
\beq \mu ~=~ 2t. \label{bound1} \eeq
This gives the boundary below the region with $\rho = 1$ for $|V| \ll t$;
this is consistent with what we see in Fig.~\ref{fig:phase_diagram}. 
Next, after doing a particle-hole transformation, $n_i \to 1 - n_i$
for all $i$, we can use arguments similar to the ones presented above 
to show that the boundary above the region $\rho=0$ in
Fig.~\ref{fig:phase_diagram_pbc} is given by
\beq \mu ~=~ - ~2t \label{bound2} \eeq
due to the appearance of one-particle states.
We would like to note here that although Eqs.~\eqref{bound1} and \eqref{bound2}
were derived assuming periodic boundary conditions, these equations will also
give the thresholds for one-hole or one-particle excitations to appear in
the bulk even when the system has open boundary conditions.

We now turn to two-hole states.
For $|V| \gg t$, we can first ignore the terms with $t$ in 
Eq.~\eqref{eq:ham}. We then see that there is a two-hole bound state
in which only the sites $i$ and $i+1$ are empty, where $i$ is odd if
$V < 0$ and is even if $V > 0$. The difference of the energy of this
state from $E_0$ is $2 \mu -|V|$ (this is less than the lowest one-hole
state which lies at $\mu - 2 t$). Hence the two-hole state has less energy
than the fully occupied state below the line $2 \mu = |V|$, i.e.,
\beq \mu ~=~ \frac{|V|}{2}. \label{bound3} \eeq
This gives the boundary below the region with $\rho = 1$ for $|V| \gg t$;
this is also consistent with Fig.~\ref{fig:phase_diagram}. (Note that 
the lowest energy of a
one-hole state is at $\mu - 2 t$ which lies above the two-hole state
at $2 \mu - |V|$ if $\mu < |V|/2$).

We will now study the form of a two-hole bound state when $|V|$ may be of 
the same order as $t$. A two-hole state will have sites $n_1$ and $n_2$ 
empty, where $n_1 < n_2$; we will denote this state as $| n_1, n_2 \ra$. 
(In this discussion, it will be convenient to
consider an infinitely large system so that we do not have to worry 
about boundary conditions). We will now consider a bound state which
has a center-of-mass momentum $k$; this lies in the range $[-2\pi,2\pi]$. 
The wave function of such a state can be written as
\beq |k \ra_B ~=~ \sum_{n_1 < n_2}~ e^{i k (n_1 + n_2)/2} ~f(n_1,n_2) 
| n_1, n_2 \ra, \eeq
where $f(n_1,n_2)$ depends on the difference $m=n_2 - n_1$ (where
$m \ge 1$) and on
whether $n_1$ is even or odd. This leads us to define
\bea f(n_1,n_2) &=& \alpha (m=n_2 - n_1) ~~{\rm if}~~ n_1 ~~{\rm is ~odd}, \nn \\
&=& \beta (m=n_2 - n_1) ~~{\rm if}~~ n_1 ~~{\rm is ~even}. \label{fn12} \eea
Then the eigenvalue equation obtained from Eq.~\eqref{eq:ham}, 
$H | k \ra_B = E_2 (k) | k \ra_B$ leads to the equations
\bea && -t ~[\alpha (m+1) e^{ik/2} ~+~ \alpha (m-1) e^{-ik/2} \nn \\
&& +~ \beta (m+1) e^{-ik/2} ~+~ \beta (m-1) e^{ik/2}] \nn \\
&& =~E_2 ~\alpha (m) ~~{\rm for}~~ m \ge 2, \nn \\
&& -t ~[\beta (m+1) e^{ik/2} ~+~ \beta (m-1) e^{-ik/2} \nn \\
&& +~ \alpha (m+1) e^{-ik/2} ~+~ \alpha (m-1) e^{ik/2}] \nn \\
&& =~ E_2 ~\beta (m) ~~{\rm for}~~ m \ge 2, \nn \\
&& -t ~[\alpha (2) e^{ik/2} ~+~ \beta (2) e^{-ik/2}] ~+~ V ~\alpha (1) ~=~ 
E_2 ~\alpha (1), \nn \\
&& -t ~[\beta (2) e^{ik/2} ~+~ \alpha (2) e^{-ik/2}] ~-~ V ~\beta (1) ~=~
E_2 ~\beta (1). \nn \\
\label{albe1} \eea

The form of Eqs.~\eqref{albe1} is invariant under the transformation
\bea \alpha (m) &\to& (-1)^m ~\beta (m), \nn \\
\beta (m) &\to& (-1)^m ~\alpha (m), \nn \\
E &\to& - ~E. \label{symm} \eea
Hence, if there is a solution of Eqs.~\eqref{albe1} with energy
$E$, there must be another solution with energy $-E$.

For every value of $k$, we can numerically solve Eqs.~\eqref{albe1} and
see if there are bound state solutions; these must satisfy 
$|\alpha (m)|$ and $|\beta (m)| \to 0$ as $m \to \infty$. If a bound state
exists, we denote its energy as $E_2 (k)$. (Then there will be another
bound state with energy $- E_2 (k)$ as discussed above).
As $|V|$ is increased from zero, we find 
that bound states first appears at $k=0$. Further, if bound states
are present for a range of values of $k$, its energy is the lowest at
$k=0$. We will therefore concentrate on the case $k=0$ henceforth.
Eqs.~\eqref{albe2} then take the form
\bea && -t ~[\alpha (m+1) ~+~ \alpha (m-1) ~+~ \beta (m+1) ~+~ \beta (m-1)] \nn \\
&& =~E_2 (0) ~\alpha (m) ~~{\rm for}~~ m \ge 2, \nn \\
&& -t ~[\beta (m+1) ~+~ \beta (m-1) ~+~ \alpha (m+1) ~+~ \alpha (m-1)] \nn \\
&& =~ E_2 (0) ~\beta (m) ~~{\rm for}~~ m \ge 2, \nn \\
&& -t ~[\alpha (2) ~+~ \beta (2)] ~+~ V ~\alpha (1) ~=~ 
E_2 (0)~\alpha (1), \nn \\
&& -t ~[\beta (2) ~+~ \alpha (2)] ~-~ V ~\beta (1) ~=~
E_2 (0) ~\beta (1). \label{albe2} \eea
These equations imply the following:
\beq \alpha (m) ~=~ \beta (m) ~~{\rm for}~~ m \ge 2, \label{albe3} \eeq
and
\bea \alpha (1) &=& -~ \frac{2 t \alpha (2)}{E_2 (0) ~-~ V}, \nn \\
\beta (1) &=& -~ \frac{2 t \alpha (2)}{E_2 (0) ~+~ V}. \label{albe4} \eea

Substituting Eq.~\eqref{albe3} in the first equation in Eq.~\eqref{albe2}
with $m \ge 3$, we obtain
\beq -2 t ~[\alpha (m+1) ~+~ \alpha (m-1)] ~=~ E_2 (0) ~\alpha (m) ~~{\rm for}
~~ m \ge 3.
\label{al1} \eeq
Ignoring an overall prefactor for $\alpha (m)$, the solution of this is 
given by 
\beq \alpha (m) ~=~ e^{- \lambda m} ~~{\rm and} ~~E_2 (0) ~=~ - 
2t ~(e^\lambda ~+~ e^{-\lambda}), \label{al2} \eeq
where $\lambda$ must be real and positive in order to have a normalizable 
state. Then the first equation in Eq.~\eqref{albe2} with $m=2$ gives
\beq \alpha (1) ~+~ \beta (1) ~=~ 2 e^{-\lambda}. \label{al3} \eeq
Combining Eqs.~\eqref{albe4}, \eqref{al2} and \eqref{al3}, we obtain
\beq e^{\lambda} ~=~ \sqrt{ \frac{V^2}{4t^2} ~-~ 1}. \label{lamb} \eeq
Hence the lowest energy of a two-hole bound state is
\beq E_2 (k=0) ~=~ -2t ~\left[ \sqrt{ \frac{V^2}{4t^2} ~-~ 1} ~+~ \frac{1}{
\sqrt{ \frac{V^2}{4t^2} ~-~ 1}} \right]. \label{e2k0} \eeq
As argued above, there must be another bound state at $k=0$ with the
energy $- E_2 (k=0)$.

Since a bound state must have $\lambda > 0$, Eq.~\eqref{lamb} implies that
a bound state appears only if $|V| > 2 \sqrt{2} t$. Since
the right hand side of Eq.~\eqref{e2k0} is always less than $-2t$,
we see that the lowest energy excitations around the fully
occupied state are two-hole bound states rather than one-hole states,
if $|V| > 2 \sqrt{2} t$.

Upon subtracting the energy of the fully occupied state, we find that the
minimum energy of a two-hole bound state is equal to $2 \mu + E_2 (k=0)$.
Hence the boundary below the region $\rho = 1$ is given by the condition
$\mu = - (1/2) E_2 (k=0)$, namely,
\beq \mu ~=~ t ~\left[ \sqrt{ \frac{V^2}{4t^2} ~-~ 1} ~+~ \frac{1}{
\sqrt{ \frac{V^2}{4t^2} ~-~ 1}} \right], \label{bound4} \eeq
provided that $|V| > 2 \sqrt{2} t$. (Note that for $|V| \gg t$, 
Eq.~\eqref{bound4} reduces to $\mu = |V| /2$ as we found earlier).

Finally, after doing a particle-hole transformation, $n_i \to 1 - n_i$, we can use similar arguments as above to
find the boundary above the region $\rho = 0$ in Fig.~\ref{fig:phase_diagram}. We find that this given by
\bea \mu &=& - ~2 t ~~{\rm for}~~ |V| ~\le~ 2 \sqrt{2} ~t, \nn \\
&=& - ~t ~\left[ \sqrt{ \frac{V^2}{4t^2} ~-~ 1} ~+~ \frac{1}{
\sqrt{ \frac{V^2}{4t^2} ~-~ 1}} \right] \nn \\
&& ~{\rm for}~~ |V| ~>~ 2 \sqrt{2} ~t. \label{bound5} \eea
Once again, we note that although Eqs.~\eqref{bound4} and \eqref{bound5} were 
derived assuming periodic conditions, they also give thresholds for two-hole
or two-particle excitations to appear in the bulk for a system with open boundary
conditions.

We will now provide an understanding of Fig.~\ref{fig:phase_diagram} obtained
for open boundary conditions. Looking closely at the boundary below the region
$\rho = 1$ for $V > 2$, we see a difference between the cases of 
periodic boundary conditions (Fig.~\ref{fig:phase_diagram_pbc}) and open boundary
conditions (Fig.~\ref{fig:phase_diagram}). This difference arises due to the
appearance of a hole state which is localized near the end of an
open system when the interaction $V_1 = V$ at the last bond (connecting sites 1 
and 2) is positive; we will call this state the hole end mode. To understand this quantitatively, we consider a state with the wave function 
\beq |\psi\rangle_B ~=~ \sum_{n=1}^\infty ~f_n ~|n \ra, \label{psiB} \eeq
where $| n \ra$ denotes a state in which only site $n$ is empty, and $n=1$
denotes the leftmost site of the system. From Eq.~\eqref{eq:ham}, we find that
the diagonal term of the Hamiltonian has a value which is $V/2$ less at
$n=1$ as compared to all other values of $n \ge 2$. Next, including the hopping 
term proportional to $-t$, we effectively obtain a tight-binding model
with a negative on-site potential $-V/2$ at site 1 and hopping amplitude $-t$
between neighboring sites $n$ and $n+1$. Eq.~\eqref{psiB} and the eigenvalue
condition $H | \psi \rangle_B = E_B | \psi \rangle_B$ lead to the equations
\bea - ~\frac{V}{2} ~f_1 ~-~ t ~f_2 &=& E_B ~f_1, \nonumber \\
- ~t ~(f_{n-1} ~+~ f_{n+1}) &=& E_B ~f_n ~~{\rm for}~~ n \ge 2. \label{fn1} \eea
The solution of the second equation in Eqs.~\eqref{fn1} is given by
\bea f_n &=& e^{-\lambda n}, \nonumber \\
E_B &=& -t ~(e^{\lambda} ~+~ e^{-\lambda}), \eea
where $\lambda$ is real and positive. Substituting this in the first equation in Eqs.~\eqref{fn1}, we obtain
\beq e^{\lambda} ~=~ \frac{V}{2t}, \eeq
which implies that the hole end mode appears only if $V > 2t$.
The energy of this mode is given by
\beq E_B ~=~ - ~\frac{V}{2} ~-~ \frac{2t^2}{V}. \eeq
This energy will be less than that of the state with all sites occupied if
the chemical potential lies below the line
\beq \mu ~=~ \frac{V}{2} ~+~ \frac{2t^2}{V}, \label{bound6} \eeq
provided that $V > 2t$. Comparing Eq.~\eqref{bound4} for a two-hole bound state
(which applies if $V > 2 \sqrt{2} t$) and Eq.~\eqref{bound4} for a hole end mode
(which applies if $V > 2 t)$, we find that $\mu$ given by Eq.~\eqref{bound6} 
always lies above $\mu$ given by Eq.~\eqref{bound4}.
Thus, if $\mu$ is gradually decreased from large values, the boundary below 
the region $\rho = 1$ is determined by the hole end mode before the two-hole bound 
states enter the picture.

Once again, we can do a particle-hole transformation and show that the boundary
above the region $\rho = 0$ in Fig.~\ref{fig:phase_diagram} for $V > 2$ is
determined by the appearance of a particle end mode. The boundary is found to 
lie at
\beq \mu ~=~ - ~\frac{V}{2} ~-~ \frac{2t^2}{V}. \label{bound7} \eeq

\section{Perturbation theory in the limit $V_1 \to - \infty$}
\label{app:appen_degenpert2}

In this Appendix we will provide a derivation of the transition line separating the phases CDW-II and PS in Fig.~\ref{fig:fullphasedia}. The analysis will proceed in the same way
as in Appendix~\ref{app:appen_degenpert1}. Given the Hamiltonian in Eq.~\eqref{eq:fermionic_ham}, we will assume that
$V_1 < 0$, with 
$- V_1 \gg t$ and $|V_2|$, and we will go up to second order in perturbation theory.
We begin by considering only the terms with $V_1$ which constitutes the
unperturbed Hamiltonian. The ground state of
this part of the Hamiltonian in Eq.~\eqref{eq:spin_ham} has a large
degeneracy: for every {\it odd} value of $i$, the 
state at sites $i$ 
and $i+1$ must be either $| 1_i 1_{i+1} \ra$
or $| 0_i 0_{i+1} \ra$, where $1_i$
and $0_i$
denote the possible fermion occupation numbers 
$n_i$ at site $i$.
We then define 
a new lattice with sites labeled as $j=(i-1)/2$ and spin-1/2 operators ${\vec \tau}_j$ which sit at the bond 
$(i,i+1)$ of the original lattice such that the states $| 1_i 1_{i+1} \ra$ and $| 0_i 0_{i+1} \ra$ correspond to $\tau_j^z = 1$ and
$\tau_j^z = -1$ respectively.

In the space of the degenerate ground states described above, we see that
the $V_2$ term in Eq.~\eqref{eq:fermionic_ham} effectively acts as $(V_2/4) 
\tau_j^z \tau_{j+1}^z$. For instance, acting
on the state $| \tau_j^z = 1, \tau_{j+1}^z =1 \ra$, the term $(V_2/4) (n_{i+1}^f - 1/2) (n_{i+2}^f - 1/2)$
gives $(V_2/4) | \tau_j^z = 1, \tau_{j+1}^z =1 \ra$.

Next, the hopping term in Eq.~\eqref{eq:fermionic_ham} of the form $-t (c_{i+1}^\dagger c_{i+2} + 
c_{i+2}^\dagger c_{i+1})$
flips the state $(n_i,n_{i+1},n_{i+2},n_{i+3})$ 
from $(1,1,0,0)$ to $(1,0,1,0)$, and vice versa.
(The first state has $\tau_j^z = 1$ and $\tau_{j+1}^z = -1$. The second state does not have a description in terms of $\tau_j^z$, and it appears as
an intermediate state in second-order perturbation theory). According to the
unperturbed Hamiltonian, the second state has an
energy which is higher than that of the first state by an amount equal to $-V_1$. As a result, second-order perturbation theory gives us a term equal to 
$(1/2) (1 - \tau_j^z \tau_{j+1}^z) (t^2 /V_1)$.

Combining the terms of order $V_2$ and $t^2 /V_1$, we
find that the low-energy effective Hamiltonian is given
by 
\beq H_{eff} ~=~ \sum_j (\frac{V_2}{4} ~-~ \frac{t^2}{2 V_1}) ~\tau_j^z \tau_{j+1}^z, \label{heff2} \eeq
up to some constants. This describes the 
classical Ising model, and it has a first-order
phase transition at $V_2 = 2 t^2/V_1$.
If $V_2 < 2 t^2/V_1$, the ground states have 
either $\tau_j = +1$ for all $j$ (so that the particle configuration is $1111 \cdots$)
or $\tau_j = -1$ for all $j$ (then the particle configuration is $0000 \cdots$).
This is the PS phase.
If $V_2 > 2 t^2/V_1$, the ground states are
$\tau_j = +1, -1, +1, -1, \cdots$ or $-1, +1, -1, +1, \cdots$, 
so that the particle configuration is either $11001100
\cdots$ or $00110011 \cdots$. This
is the CDW-II phase. The transition between the two phases
is first-order because if $E_0 (V_1, V_2)$ denotes the
ground state energy and we hold $V_1$ fixed, then the
derivative $dE_0 /dV_2$ has a discontinuity when we cross
the line $V_2 = 2 t^2/V_1$.

\bibliography{ladder.bib}

\end{document}